\documentclass[a4paper,11pt,preprint]{article}
\pdfoutput=1 

\usepackage{jheppub} 

\usepackage[T1]{fontenc} 

\usepackage{graphicx}
\usepackage{amsmath}
\usepackage{amssymb}

\def\ba{\begin{eqnarray}}
\def\ea{\end{eqnarray}}

\def\be{\begin{equation}}
\def\ee{\end{equation}}

\newcommand*\barM{{\overline{\cal M}\hspace{0.5mm}}}

\title{The non-equilibrium attractor for kinetic theory in relaxation time approximation}

\author{M. Strickland}
\affiliation{Department of Physics, Kent State University, Kent, OH 44242 United States}

\emailAdd{mstrick6@kent.edu}

\abstract{
I demonstrate that the concept of a non-equilibrium attractor can be extended beyond the lowest-order moments typically considered in hydrodynamic treatments.  Using a previously obtained exact solution to the relaxation-time approximation Boltzmann equation for a transversally homogeneous and boost-invariant system subject to Bjorken flow, I derive an equation obeyed by all moments of the one-particle distribution function.  Using numerical solutions, I show that, similar to the pressure anisotropy, all moments of the distribution function exhibit attractor-like behavior wherein all initial conditions converge to a universal solution after a short time with the exception of moments which are sensitive to modes with zero longitudinal momentum and high transverse momentum.  In addition, I compute the exact solution for the distribution function itself on very fine lattices in momentum space and demonstrate that (a) an attractor for the full distribution function exists and (b) solutions with generic initial conditions relax to this solution, first at low momentum and later at high momentum.
}

\keywords{quark-gluon plasma, relativistic heavy-ion collisions, relativistic kinetic theory, relativistic dissipative hydrodynamics}

\begin{document} 
\maketitle
\flushbottom

\section{Introduction}
\label{sect:intro}

Understanding the thermalization of systems which are driven far from equilibrium is of fundamental importance to many topical areas in physics.  In the context of ultrarelativistic heavy-ion collisions (URHICs), for example, one would like to understand how quickly the quark-gluon plasma (QGP) approaches thermal equilibrium.  In URHICs, a significant impediment to fast thermalization is the rapid longitudinal expansion of the QGP which induces large violations of local-rest-frame (LRF) momentum-isotropy at early times in both the weak \cite{Baier:2000sb,Blaizot:2001nr,Romatschke:2003ms,Arnold:2003rq,Arnold:2004ti,Mrowczynski:2004kv,Rebhan:2004ur,Rebhan:2005re,Romatschke:2005pm,Romatschke:2006nk,Romatschke:2006wg,Rebhan:2008uj,Fukushima:2011nq,Kurkela:2011ti,Kurkela:2011ub,Blaizot:2011xf,Attems:2012js,Berges:2012iw,Gelis:2013rba} and strong coupling limits \cite{Chesler:2008hg,Beuf:2009cx,Chesler:2009cy,Heller:2011ju,Heller:2012je,Heller:2012km,vanderSchee:2012qj,Casalderrey-Solana:2013aba,Heller:2013oxa,Keegan:2015avk,Chesler:2015bba,Kurkela:2015qoa,Chesler:2016ceu,Attems:2016ugt,Attems:2016tby,Attems:2017zam}.  These LRF momentum-space anisotropies diminish as a function of time, but persist throughout the entire QGP lifetime \cite{Strickland:2013uga}.  Despite this, it has recently been found that, in a variety of settings including AdS/CFT simulations of non-equilibrium dynamics, simple kinetic models, and QCD-based kinetic approaches  \cite{Noronha:2011fi,Heller:2015dha,Keegan:2015avk,Florkowski:2017olj,Romatschke:2017vte,Bemfica:2017wps,Spalinski:2017mel,Romatschke:2017acs,Behtash:2017wqg,Florkowski:2017jnz,Florkowski:2017ovw,Strickland:2017kux,Almaalol:2018ynx,Denicol:2018pak,Behtash:2018moe}, the dynamics possesses a ``non-equilibrium attractor'' which can be well-described by relativistic dissipative hydrodynamics after a relatively short amount of time in the center of the fireball ($\lesssim 1$ fm/c).\footnote{For a recent review see Ref.~\cite{Florkowski:2017olj}.}  This process has been dubbed ``hydrodynamization'' because, although evolution in each case is well-described by dissipative hydrodynamics, the corrections to thermal equilibrium are large with the largest violations reflected in the strong early-time pressure anisotropies present in the system.  These studies provide insight into the degree to which one can apply traditional hydrodynamic approaches in non-equilibrium settings and, at the same time, provide insight into the magnitude of non-hydrodynamic modes' impact on the dynamics.  

The hydrodynamization of the QGP can be understood by studying the approach of different initial conditions to a non-equilibrium attractor \cite{Heller:2015dha}.  For example, one finds in practice that the pressure anisotropy ($P_L/P_T$) developed from a set of different initial conditions all approach a universal attractor solution in a time on the order of a few fm/c with the time scale increasing as one decreases the effective coupling or initial temperature.  Once the solutions have collapsed onto the attractor, information about the details of the initial condition used are wiped out, suggesting a kind of pseudo-thermalization of the system \cite{Strickland:2017kux}.  One shortcoming of prior works is that, when discussing the attractor, past authors (including this author) have restricted their attention to the amplitude, $\varphi$, which can be related to a particular ratio of a linear combination of low-order moments of the one-particle distribution function.  A natural question is whether or not attractor-like behavior is observed in the evolution of all moments of the one-particle distribution function and, if an attractor for higher-moments exists, what is the timescale for a generic initial condition to approach the attractor for a given moment.

In this paper, I address this question by considering the behavior of general moments of the one-particle distribution function in the case of a 0+1d conformal system subject to boost-invariant longitudinal Bjorken flow and a relaxation-time approximation (RTA) collisional kernel.  For this purpose, I make use of a semi-analytic exact solution of the conformal RTA Boltzmann equation which has been obtained previously \cite{Florkowski:2013lza,Florkowski:2013lya}.  This exact solution takes the form of a one-dimensional integral equation for the energy density which can be solved iteratively and mapped to a solution for the effective temperature of the conformal system.  Once the effective temperature is known, one can solve for all moments of the one-particle distribution function and the distribution function itself.  I will present results for a subset of moments which are indicative of the general behavior seen.  

I find that there exists attractor-like behavior in all moments, however, low-order moments which contain only powers $E$ and not $p_z$ exhibit a slower approach to their respective attractors.  This is explained in terms of a spectrum of modes with nearly zero $p_z$ and an average $p_T$ which increases with time.  These modes have a small impact on the overall dynamics but dominate these particular moments.  For all other moments, I find that the timescale for approach to the attractor only weakly depends on the order of the moment considered.  The convergence of all modes to their respective attractor on approximately the same time scale suggests that the evolution of the full one-particle distribution function exhibits attractor-like behavior.  I demonstrate that it does by solving for the full one-particle distribution function associated with the attractor and then I study the flow of the distribution function towards the distribution function's attractor.

The structure of this paper is as follows.  In Sec.~\ref{sect:setup}, I review the basic setup for finding the exact solution to the 0+1d RTA Boltzmann equation.  In Sec.~\ref{sect:genmomevol}, I present the integral equation obeyed by a general moment ${\cal M}^{nm}$ of the one-particle distribution function and demonstrate that, for low-order moments, it reduces to results obtained previously in the literature.  In Sec.~\ref{sect:dissipativehydro}, I calculate the moments using anisotropic hydrodynamics and second-order viscous hydrodynamics.  In Sec.~\ref{sect:results}, I present my numerical results and discuss.  In Sec.~\ref{sect:conclusions}, I present my conclusions and an outlook for the future. 

\section{Setup and review}
\label{sect:setup}

In this section I review how to obtain the exact solutions to the 0+1d RTA Boltzmann equation.  The method is based on the original work presented in Refs.~\cite{Florkowski:2013lza,Florkowski:2013lya}.\footnote{For related works which extend this solution to non-conformal systems, quantum statistics, Gubser flow, and coupled quark-gluon RTA kinetic equations, I refer the reader to Refs.~\cite{Florkowski:2014sfa,Florkowski:2014sda,Denicol:2014xca,Denicol:2014tha,Florkowski:2017jnz,Maksymiuk:2017cnv}.}

\subsection{RTA Boltzmann equation}
\label{sect:rta}

My starting point is the RTA Boltzmann equation
\begin{equation}
 p^\mu \partial_\mu  f(x,p) =  C[ f(x,p)] , 
\label{kineq}
\end{equation}
with
\begin{eqnarray}
C[f] = \frac{p \cdot u}{\tau_{\rm eq}} \left( f_{\rm eq}-f \right) .
\label{col-term}
\end{eqnarray}
The quantity $\tau_{\rm eq} = 5\bar\eta/T$ is the relaxation time where $\bar\eta=\eta/s$ is the shear viscosity to entropy density ratio and $T$ is the local spacetime-dependent effective temperature, which is proportional to the fourth root of the local energy density.  For a conformal system, the equilibrium distribution function $f_{\rm eq}$ may be taken to be a Bose-Einstein, Fermi-Dirac, or Boltzmann distribution.    Herein, I will assume that $f$ is given by a Boltzmann distribution
\begin{eqnarray}
f_{\rm eq} = \exp\left(- \frac{p \cdot u}{T} \right) .
\label{Boltzmann}
\end{eqnarray}
The effective temperature $T$ can be obtained via the Landau matching condition which demands that the energy density calculated from the distribution function $f$ is equal to the energy density determined from an equilibrium distribution, $f_{\rm eq}$.  The quantity $u^\mu$ is the four-velocity of the local rest frame of the matter (fluid four velocity).  I will assume Bjorken flow, in which case the Minkowski-space components of the four-flow are \mbox{$u^\mu = (t/\tau,0,0,z/\tau)$}, where $\tau = \sqrt{t^2-z^2}$ is the longitudinal proper-time.  In Milne coordinates, Bjorken flow is static, i.e. $u^\tau=1$ and $u^{x,y,\varsigma}=0$.

The use of this simple form of the kinetic equation given by Eqs.~\eqref{kineq} and \eqref{col-term} is motivated by the fact that there are many results obtained within this approximation, allowing one to make comparisons with other approaches.  In particular, there exist consistent second and third-order calculations of the kinetic coefficients in RTA, for example, see Refs.~\cite{Denicol:2010xn,Denicol:2011fa,Jaiswal:2013npa,Jaiswal:2013vta,Denicol:2014mca,Florkowski:2015lra}.  Finally, and perhaps most importantly, in this simple case it is possible to solve the kinetic equation exactly using straightforward numerical algorithms.

\subsection{Thermodynamic functions}
\label{sect:thermo}

For a single species of scalar massless particles obeying equilibrium classical statistics the particle density, entropy density, energy density, and pressure are
\begin{eqnarray}
n_{\rm eq} = \frac{T^3}{\pi^2}, \quad
s_{\rm eq} = \frac{4 T^3}{\pi^2},
\nonumber \\
\varepsilon_{\rm eq} = \frac{3 T^4}{\pi^2}, \quad
P_{\rm eq} = \frac{T^4}{\pi^2},
\label{eq-therm}
\end{eqnarray}
In what follows I make use of the relation $\varepsilon_{\rm eq} = 3 P_{\rm eq}$ when a specification of the equilibrium equation of state is required.

\subsection{Boost-invariant variables}
\label{sect:boostinvvar}

In the case of one-dimensional boost-invariant expansion (0+1d), all scalar functions of spacetime depend only on the longitudinal proper time $\tau$.  To proceed, in addition to the timelike flow four-vector $u^\mu$ one can introduce a spacelike vector that is orthogonal in all frames and corresponds to the z-direction in the local rest frame of the matter,  $z^\mu_{\rm LAB} = \left(z/\tau,0,0,t/\tau\right)$ and $z^\mu_{\rm LRF} = (0,0,0,1)$.

The one-particle distribution function $f(x,p)$ is a scalar under Lorentz transformations.  The requirement of boost invariance implies that in this case $f(x,p)$ may depend only on three variables: $\tau$, $w$, and $\vec{p}_T$ \cite{Bialas:1984wv,Bialas:1987en}.  The boost-invariant variable $w$ is defined by 
\begin{equation}
w =  tp_L - z E \, .
\label{w}
\end{equation}
Note that, in the prior equation, $z$ is the spatial coordinate, not to be confused with the basis vector $z^\mu$.
With the help of $w$ and $\vec{p}_T$ one defines
\begin{equation}
v(\tau,w,p_T) = Et-p_L z = 
\sqrt{w^2+\left( m^2+\vec{p}_T^{\,\,2}\right) \tau^2} \, .  
\label{v}
\end{equation}
Using (\ref{w}) and (\ref{v}) one can easily find the energy and the longitudinal momentum of a particle 
\begin{equation}
E= p^0 = \frac{vt+wz}{\tau^2} \, ,\quad p_L=\frac{wt+vz}{\tau^2} \, .  
\label{p0p3}
\end{equation}
The momentum integration measure is 
\begin{equation}
dP =   \frac{d^4p}{(2\pi)^4} \, 2\pi \delta \left( p^2-m^2\right) 2 \theta (p^0)
=\frac{dp_L}{(2\pi)^3p^0}d^2p_T =\frac{dw \, d^2p_T }{(2\pi)^3v}\, .  
\label{dP}
\end{equation}
In the following I will consider massless partons, $m=0$.

\subsubsection*{Boost-invariant form of the kinetic equation}
\label{sect:binvkineq}

Using the boost-invariant variables introduced in the previous section, one finds~\cite{Florkowski:2013lza,Florkowski:2013lya}
\begin{eqnarray}
p^\mu \partial_\mu f = 
\frac{v}{\tau} \frac{\partial f}{\partial \tau}, \quad 
p_\mu u^\mu = \frac{v}{\tau} \,  \quad 
p_\mu z^\mu = - \frac{w}{\tau} \, .
\label{binvterms}
\end{eqnarray}
Using Eqs.~(\ref{binvterms}) in Eq.~(\ref{kineq}) and simplifying, the 0+1d Boltzmann equation takes a particularly simple form~\cite{Florkowski:2013lza,Florkowski:2013lya}
\begin{eqnarray}
\frac{\partial f}{\partial \tau}  &=& 
\frac{f_{\rm eq}-f}{\tau_{\rm eq}} \, ,
\end{eqnarray} 
where the equilibrium distribution function can be written as
\begin{eqnarray}
f_{\rm eq}(\tau,w,p_T) =
\exp\!\left[
- \frac{\sqrt{w^2+p_T^2 \tau^2}}{T(\tau) \tau}  \right].
\label{eqdistform}
\end{eqnarray}
In the results section, I make use of the fact that symmetries require that $f(\tau,w,\vec{p}_T)$ be an even function of $w$ and that it depends only on the magnitude of the transverse momentum $\vec{p}_T$.

\subsection{Exact solution}
\label{sect:exactsol}

Once cast in the form \eqref{eqdistform}, the solution becomes straightforward, with the result being~\cite{Baym:1984np,Florkowski:2013lza,Florkowski:2013lya}
\be
f(\tau,w,p_T) = D(\tau,\tau_0) f_0(w,p_T) 
+  \int_{\tau_0}^\tau \frac{d\tau^\prime}{\tau_{\rm eq}(\tau^\prime)} \, D(\tau,\tau^\prime) \, 
f_{\rm eq}(\tau^\prime,w,p_T) \, ,  
\label{eq:exactsolf}
\ee
where $D$ is the damping function
\begin{eqnarray}
D(\tau_2,\tau_1) = \exp\left[-\int\limits_{\tau_1}^{\tau_2}
\frac{d\tau^{\prime\prime}}{\tau_{\rm eq}(\tau^{\prime\prime})} \right] .
\end{eqnarray}
At \mbox{$\tau=\tau_0$} the distribution function $f$ reduces to the initial distribution function, $f_0$.  For the conformal RTA Boltzmann equation, one has~\cite{Denicol:2010xn,Denicol:2011fa}
\begin{eqnarray}
\tau_{\rm eq}(\tau) = \frac{5 {\bar \eta}}{T(\tau)},
\label{taueq}
\end{eqnarray}
where ${\bar \eta} \equiv \eta/s$ is the ratio of the shear viscosity to entropy density.

In order to solve for $f$ using Eq.~\eqref{eq:exactsolf} one needs to know $T(\tau)$.  An efficient way to do this is by considering the integral equation for the energy density obtained by integrating both sides of Eq.~\eqref{eq:exactsolf} times $(p \cdot u)^2$, i.e. $\varepsilon = \int dP \, (p \cdot u)^2 f$.  After performing this operation on Eq.~\eqref{eq:exactsolf}, the left hand side becomes the non-equilibrium energy density, $\varepsilon(\tau)$, while the first term on the right hand side encodes the free-streaming contribution to the energy density evolution, and the second term on the right hand side dominates the late-time behavior for any $\tau_{\rm eq} \neq \infty$.   To close the equation, one implements the Landau matching condition $\varepsilon(\tau) = \varepsilon_{\rm eq}(T)$ on the left-hand-side so that one obtains an integral equation for $\varepsilon_{\rm eq}(T) \propto T^4(\tau)$.  Before presenting this, however, one needs to specify the form of the initial distribution function $f_0$ appearing in Eq.~\eqref{eq:exactsolf} in order to fix the free streaming contribution.

\subsubsection*{Initial distribution function}
\label{sect:initdistr}

In this paper, I consider initial conditions of Romatschke-Strickland form \cite{Romatschke:2003ms} with a Boltzmann distribution as the underlying isotropic distribution 
\begin{eqnarray}
f_0(w,p_T) &=& 
\exp\left[
-\frac{\sqrt{(p\cdot u)^2 + \xi_0 (p\cdot z)^2}}{\Lambda_0} \, \right] \nonumber \\
&=& 
\exp\left[
-\frac{\sqrt{(1+\xi_0) w^2 + p_T^2 \tau_0^2}}{\Lambda_0 \tau_0}\, \right].
\label{RS}
\end{eqnarray}
This reduces to an isotropic Boltzmann distribution if the anisotropy parameter $\xi_0=\xi(\tau_0)$ vanishes. In this case, the transverse momentum scale $\Lambda_0$ is equal to the system's initial temperature $T_0$. By direct calculation one obtains
\begin{eqnarray}
\varepsilon^0(\tau) &=& \int dP \, (p\cdot u)^2 f_0(w,p_T)  = 
\frac{3 T_0^4}{\pi^2} \, 
\frac{{\cal H}\hspace{-1mm}
\left( \frac{\alpha_0\tau_0 }{\tau} \right)}
{{\cal H}\hspace{-1mm}\left(\alpha_0\right)} \, ,
\label{eps0tauRS}
\end{eqnarray}
where 
\be
{\cal H}(y) = y \int\limits_0^\pi d\phi \, \sin\phi \, \sqrt{y^2 \cos^2\phi+\sin^2\phi} \, ,
\ee
with
\begin{eqnarray}
\alpha(\tau) = (1+\xi(\tau))^{-1/2} \, ,
\label{iks}
\end{eqnarray}
and $\alpha_0=\alpha(\tau_0)$ denotes the initial value of the anisotropy parameter with $0 \leq \alpha \leq \infty$.

\subsubsection*{Integral equation for the effective temperature}

The resulting integral equation for the effective temperature is~\cite{Florkowski:2013lza,Florkowski:2013lya}
\be
T^4(\tau) = D(\tau,\tau_0)T_0^4 \, 
\frac{{\cal H}\hspace{-1mm}
\left( \frac{\alpha_0\tau_0 }{\tau} \right)}
{{\cal H}\hspace{-1mm}\left(\alpha_0\right)} + \int_{\tau_0}^\tau \frac{d\tau^\prime}{2 \tau_{\rm eq}(\tau^\prime)} \, D(\tau,\tau^\prime) \, 
T^4(\tau^\prime) {\cal H}\!\left(\frac{\tau^\prime}{\tau}\right).
 \label{solT1}  
\ee
This equation can be solved iteratively by constructing a discrete lattice in proper-time, making an initial guess for $T(\tau_i)$, and using numerical quadratures to compute the required integrals on the right hand side.  I will provide details of the lattice size, spacing, tolerance, etc. in the results section.  Once the solution for $T(\tau)$ is obtained by iterating this procedure, it can then be used in Eq.~\eqref{eq:exactsolf} to reconstruct the full one-particle distribution function point-by-point in momentum space.  In principle, one could compute $f$ and then determine moments from that, however, this proves to be quite inefficient and computationally demanding.  Instead, one can derive integral equations which can be used to determine the evolution of all moments of the distribution function.  

\section{Evolution equation for a general moment}
\label{sect:genmomevol}

To extend the treatment presented in the previous section to general moments, one introduces 
\be
{\cal M}^{nm}[f] \equiv \int dP \,(p \cdot u)^n \, (p \cdot z)^{2m} \, f(\tau,w,p_T) \, .
\label{eq:genmom1}
\ee
In principle, powers of $p_T^{2\ell}$ could also appear in a general moment, however, such moments can be expressed as a linear combination of the two-index moment appearing above using $p^2=0$ to write $p_T^{2\ell} = [(p \cdot u)^2 - (p\cdot z)^2]^\ell$.

For $n=1$ and $m=0$ one obtains the number density
\be
n = {\cal M}^{10} = \int dP \, (p \cdot u) \, f(\tau,w,p_T)   = j^0_{\rm LRF} \, .
\ee
For $n=2$ and $m=0$ one obtains the energy density
\be
\varepsilon = {\cal M}^{20} = \int dP \, (p \cdot u)^2 \, f(\tau,w,p_T)  = T^{00}_{\rm LRF} \, ,
\ee
and for $n=0$ and $m=1$ one obtains the longitudinal pressure
\be
P_L = {\cal M}^{01} = \int dP \, (p \cdot z)^2 \, f(\tau,w,p_T)  = T^{zz}_{\rm LRF} \, .
\ee
For a conformal system, one can use $\varepsilon = 2 P_T + P_L$ to determine the transverse pressure.  This follows by using the mass shell condition, $p_T^{2\ell} = [(p \cdot u)^2 - (p\cdot z)^2]^\ell$.

In the general case, using the boost-invariant variables introduced earlier, the moment integral necessary becomes
\ba
{\cal M}^{nm}[f] &=& \int  \frac{dw \, d^2p_T }{(2\pi)^3v} \left( \frac{v}{\tau} \right)^n \left( \frac{w}{\tau} \right)^{2m} \, f(\tau,w,p_T) \, , \nonumber \\
&=& \frac{1}{(2\pi)^3 \, \tau^{n+2m}} \int  dw \, d^2p_T  \, v^{n-1} w^{2m} \, f(\tau,w,p_T)
\ea
Taking a general moment of Eq.~(\ref{eq:exactsolf}) one obtains
\be
{\cal M}^{nm}(\tau) = D(\tau,\tau_0) {\cal M}^{nm}_0(\tau)  + \int_{\tau_0}^\tau \frac{d\tau^\prime}{\tau_{\rm eq}(\tau^\prime)} \, D(\tau,\tau^\prime) \, 
{\cal M}^{nm}_{\rm eq}(\tau') .  \nonumber
\ee
Using the results obtained above, one can write this as
\ba
{\cal M}^{nm}(\tau) &=& \frac{\Gamma(n+2m+2)}{(2\pi)^2} \Bigg[ D(\tau,\tau_0) 2^{(n+2m+2)/4} T_0^{n+2m+2} \frac{{\cal H}^{nm}\!\left( \frac{\alpha_0 \tau_0}{\tau} \right)}{[{\cal H}^{20}(\alpha_0)]^{(n+2m+2)/4}} \nonumber \\
&& + \int_{\tau_0}^\tau \frac{d\tau^\prime}{\tau_{\rm eq}(\tau^\prime)} \, D(\tau,\tau^\prime) \, 
T^{n+2m+2}(\tau') {\cal H}^{nm} \hspace{-1mm} \left( \frac{\tau'}{\tau} \right) \Bigg] ,
\label{eq:meqfinal}
\ea
with
\be
{\cal H}^{nm}(y) = \tfrac{2y^{2m+1}}{2m+1}  {}_2F_1(\tfrac{1}{2}+m,\tfrac{1-n}{2};\tfrac{3}{2}+m;1-y^2)  \, .
\ee
Eq.~\eqref{eq:meqfinal} is one of the main results obtained herein.

\subsection*{Cross checks}

As a check of the general moment equation, one can verify that it reduces to known equations for the low-order moments of the distribution function available in the literature.  Taking the $n=2$ and $m=0$, and relabeling ${\cal H}^{20} \rightarrow {\cal H}$, one obtains
\be
{\cal M}^{20}(\tau) = \frac{3}{2\pi^2} \Bigg[ 2 D(\tau,\tau_0) T_0^4 \frac{{\cal H}\!\left( \frac{\alpha_0 \tau_0}{\tau} \right)}{{\cal H}(\alpha_0)} \nonumber \\
+ \int_{\tau_0}^\tau \frac{d\tau^\prime}{\tau_{\rm eq}(\tau^\prime)} \, D(\tau,\tau^\prime) \, 
T^4(\tau') {\cal H}\hspace{-1mm} \left( \frac{\tau'}{\tau} \right) \Bigg] .  
\ee
Using the fact that ${\cal M}^{20}(\tau) = \varepsilon(\tau) = \varepsilon_{\rm eq}(\tau) = 3 T^4(\tau)/\pi^2$, this reduces to the known integral equation for the effective temperature~\cite{Florkowski:2013lza,Florkowski:2013lya}
\be
T^4(\tau) = D(\tau,\tau_0) T_0^4 \frac{{\cal H}\!\left( \frac{\alpha_0 \tau_0}{\tau} \right)}{{\cal H}(\alpha_0)} \nonumber \\
+ \int_{\tau_0}^\tau \frac{d\tau^\prime}{2 \tau_{\rm eq}(\tau^\prime)} \, D(\tau,\tau^\prime) \, 
T^4(\tau') {\cal H}\hspace{-1mm} \left( \frac{\tau'}{\tau} \right)  .  \nonumber
\label{t4eq}
\ee

Taking $n=0$ and $m=1$, and relabeling ${\cal H}^{01} \rightarrow {\cal H}_L$, one obtains the longitudinal pressure~\cite{Florkowski:2013lza,Florkowski:2013lya}
\be
P_L(\tau) = {\cal M}^{01}(\tau) = \frac{3}{\pi^2} \Bigg[ D(\tau,\tau_0) T_0^4 \frac{{\cal H}_L\!\left( \frac{\alpha_0 \tau_0}{\tau} \right)}{{\cal H}(\alpha_0)} 
+ \int_{\tau_0}^\tau \frac{d\tau^\prime}{2\tau_{\rm eq}(\tau^\prime)} \, D(\tau,\tau^\prime) \, 
T^4(\tau') {\cal H}_L \hspace{-1mm} \left( \frac{\tau'}{\tau} \right) \Bigg] .  \nonumber
\ee

\section{Evolution of the moments in dissipative hydrodynamics}
\label{sect:dissipativehydro}

In the results section I will compare results obtained using Eq.~\eqref{eq:meqfinal} with those obtained using anisotropic hydrodynamics (aHydro) \cite{Florkowski:2010cf,Martinez:2010sc,Tinti:2013vba,Alqahtani:2017mhy} and second-order viscous hydrodynamics (vHydro) \cite{Denicol:2010xn,Denicol:2011fa}.  In anticipation of this, I now list the results for a general moment in both schemes.  For aHydro the moments depend only on the anisotropy parameter $\alpha$ and for vHydro they depend only on the ratio of the shear viscous correction $\pi = \pi^\varsigma_\varsigma$ to the energy density.

\subsection{Evolution of the moments in anisotropic hydrodynamics}

In aHydro, one uses the first and second moments of the Boltzmann equation to solve for the evolution of $\alpha$ and $\Lambda$ with the one-particle distribution function assumed to be of the form~\cite{Florkowski:2010cf,Martinez:2010sc}
\begin{eqnarray}
f_{\rm aHydro}(w,p_T) &=& 
\exp\left[
-\frac{\sqrt{(p\cdot u)^2 + \xi(\tau) (p\cdot z)^2}}{\Lambda(\tau)} \, \right] \nonumber \\
&=& 
\exp\left[
-\frac{\sqrt{[1+\xi(\tau)] w^2 + p_T^2 \tau^2}}{\Lambda(\tau) \tau}\, \right].
\label{RS2}
\end{eqnarray}
Using this form, one finds 
\be
{\cal M}^{nm}_{\rm aHydro}(\tau) = \frac{\Gamma(n+2m+2) \Lambda^{n+2m+2}(\tau)}{(2\pi)^2} {\cal H}^{nm}\!\left( \alpha(\tau) \right) ,
\label{mres_ahydro}
\ee
with $\alpha(\tau) = 1/\sqrt{1+\xi(\tau)}$.

When comparing results it is useful to rescale each moment by its equilibrium value
\be
{\cal M}^{nm}_{\rm eq}(\tau) = \frac{\Gamma(n+2m+2) T^{n+2m+2}(\tau)}{2\pi^2 (2m +1)} \, .
\label{meqres2}
\ee
For aHydro, this gives
\be
\barM^{nm}_{\rm aHydro}(\tau) = 2^{(n+2m-2)/4} (2m+1) \frac{{\cal H}^{nm}(\alpha)}{[{\cal H}^{20}(\alpha)]^{(n+2m+2)/4}} \, .
\label{eq:ahydromoms}
\ee
where I have introduced the scaled moments
\be
\barM^{nm}(\tau) \equiv \frac{{\cal M}^{nm}(\tau)}{{\cal M}^{nm}_{\rm eq}(\tau)} \, .
\label{mbardef}
\ee
It is straightforward to verify that in the isotropic (equilibrium) limit, one has \mbox{$\barM^{nm}_{\rm aHydro}(\tau) =1$}.

\subsection{Evolution of the moments in second-order viscous hydrodynamics}

In vHydro, the 0+1d one-particle distribution function takes the form \cite{Denicol:2010xn,Denicol:2011fa}
\be
f = f_{\rm eq}(T(\tau)) \left[ 1 + \frac{3 \bar\pi(\tau)}{16 T^2(\tau)} \left\{ (p\cdot u)^2 - 3 (p\cdot z)^2 \right\} \right] ,
\ee
where $\bar\pi = \pi/\varepsilon$.  Computing ${\cal M}^{nm}$ using this form, one finds
\be
{\cal M}^{nm}_{\rm vHydro}(\tau) = {\cal M}_{\rm eq}^{nm}(\tau) + \frac{3 \bar\pi(\tau)}{16 T^2(\tau)} \left[  {\cal M}_{\rm eq}^{n+2,m}(\tau) - 3 {\cal M}_{\rm eq}^{n,m+1}(\tau) \right] ,
\ee
with $\bar\pi \equiv \pi/\varepsilon$.   Computing the scaled moment, one obtains
\be
\barM^{nm}_{\rm vHydro}(\tau) = 1 - \frac{3 m (n+2m+2)(n+2m+3)}{4(2m+3)} \bar\pi \, .
\label{eq:vhydromoms}
\ee
Note that the Navier-Stokes (NS) result for a general moment can be obtained from the above expression by taking $\bar\pi = 16 \bar\eta/(9 \tau T)$.  Also note that from the above equation one finds that $\barM^{n0}_{\rm vHydro}(\tau) = 1$.

\subsection*{Navier-Stokes thermalization time}

Before proceeding, I mention that using the above relation and taking the Navier-Stokes limit, one can solve for the scaled time $\overline{w} \equiv \tau/\tau_{\rm eq}$ where $\barM^{nm}_{\rm NS}(\tau) = 1 - \delta$ which gives.
\be
\overline{w}_{\rm therm}^{\rm NS}  =  \frac{12}{45}  \frac{1}{\delta}  \frac{m (n+2m+2)(n+2m+3)}{(2m+3)} \, .
\ee
As one can see from this result, the Navier-Stokes limit predicts that large $m$ and $n$ moments will equilibrate at a later time than small $m$ and $n$ moments.  In the results section, I will compare this result to the thermalization time extracted from the exact solution to the Boltzmann equation.\footnote{As defined above, at the thermalization time, the system still possesses (small) non-equilibrium corrections which induce, e.g., pressure anisotropy and deviations of all moments from the true $\tau \rightarrow \infty$ equilibrium solution.}

\begin{figure*}[t!]
\centerline{
\includegraphics[width=1\linewidth]{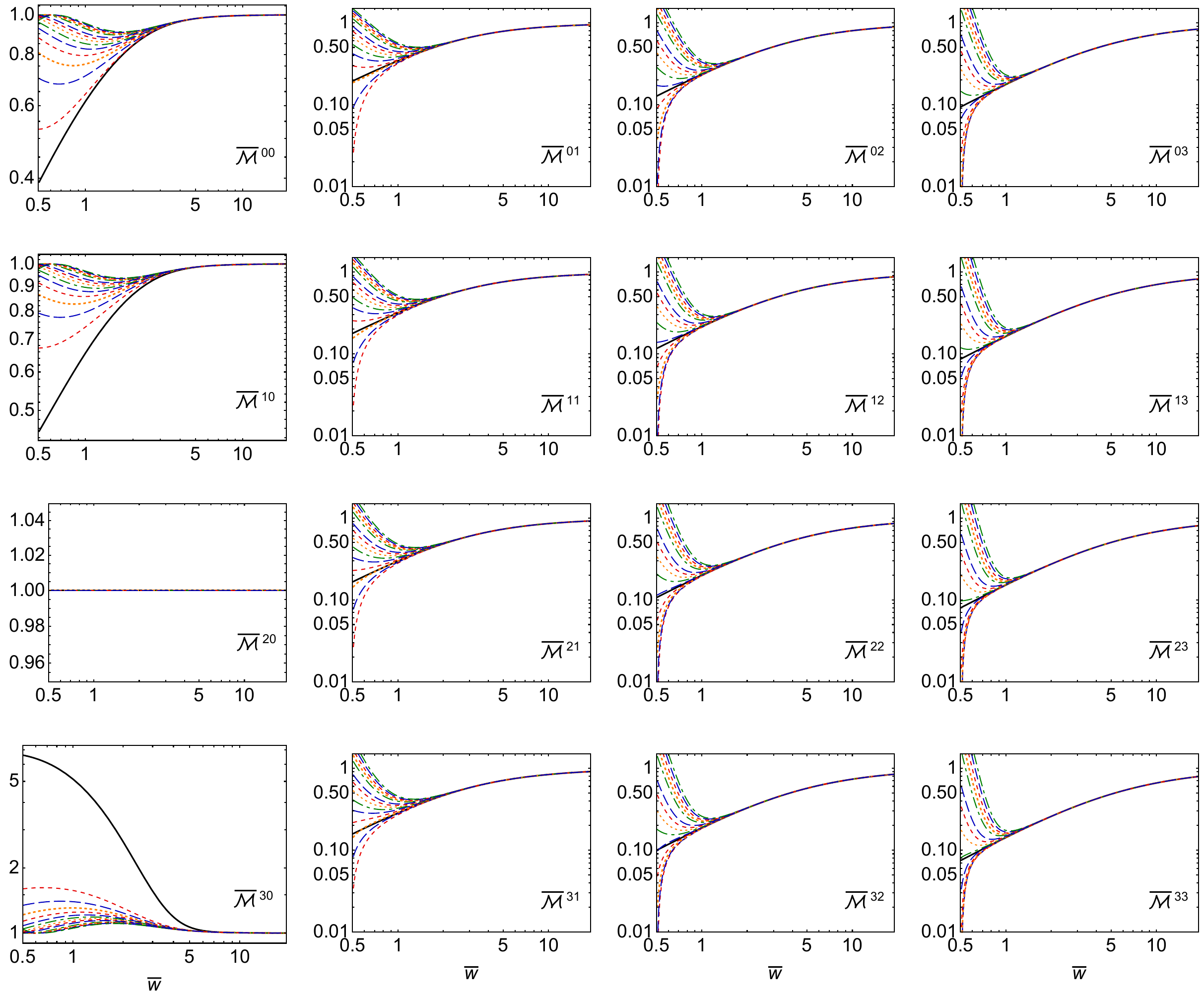}
}
\caption{Scaled moments $\barM^{nm}$ obtained from the exact attractor solution (solid black line) compared to a set of exact solutions (various colored dotted and dashed lines) initialized at $\tau = 0.1$ fm/c with varying initial pressure anisotropy.  The horizontal axis is $\overline{w} \equiv \tau/\tau_{\rm eq} = \tau T/5 \bar\eta$.  Panels show a grid in $n$ and $m$. }
\label{fig:attractorGridSols}
\end{figure*}
 
\section{Numerical results}
\label{sect:results}

I will now present some representative numerical solutions using different initial conditions and the attractor solution to which they flow.  For a given value of $\bar\eta$ and set of initial conditions specified by $\alpha_0$ and $T_0$, I solve the integral equation (\ref{solT1}) numerically.  For this purpose I wrote a CUDA-based GPU code which allows one to efficiently solve the integral equation \eqref{solT1} efficiently on very large lattices.  The code uses a logarithmically-spaced grid in proper-time in order to more accurately account for the effects of large early time gradients.  The code is included in the arXiv bundle for this paper and is also publicly available for download using the link provided in Ref.~\cite{MikeCodeDB}.  Once the solution for the effective temperature is obtained, one can use Eq.~\eqref{eq:meqfinal} to obtain any moment required.  One can also use Eq.~\eqref{eq:exactsolf} to reconstruct the full one-particle distribution in a grid in momentum-space.  For all results presented here, I iterated the integral equation for the temperature  (\ref{solT1}) until the result converged to sixteen digits at all values of $\tau$.

I analyze the (pseudo-)thermalization of the system by considering the scaled moments \eqref{meqres2} as a function of the scaled time $\overline{w} \equiv \tau/\tau_{\rm eq} = \tau T/5 \bar\eta$.\footnote{The scaled time $\overline{w}$ should not be confused with the boost-invariant variable $w$ introduced earlier.}$^,$\footnote{The definition of $\overline{w}$ used herein is the same (up to a constant) as the scaled proper time introduced in Ref.~\cite{Heller:2016rtz}.  In both cases, one is essentially dividing the proper time by the microscopic relaxation time.}  The rate at which the scaled moments approach unity provides information about the thermalization of the system.\footnote{One exception is $\barM^{20}$ which equals unity at all times due to energy conservation.} Some familiar quantities like the scaled number density and longitudinal pressure are given by $\barM^{10}(\tau) = n(\tau)/n_{\rm eq}(\tau)$ and $\barM^{01}(\tau) = P_L(\tau)/P_{\rm eq}(\tau)$, respectively.

In Fig.~\ref{fig:attractorGridSols}, I compare the attractor (black solid line) with a set of representative solutions (dashed/dotted colored lines) with differing levels of initial momentum-space anisotropy ($0.1 \leq \alpha_0 \leq 1.5$) and fixed initial temperature $T_0 = 1$ GeV at $\tau_0 = 0.1$ fm/c.  For the solutions with different initial conditions (dashed/dotted colored lines) I used 2048 points spaced logarithmically between $\tau = 0.1$ and 100 fm/c.  For the attractor solution, I used 4096 points spaced logarithmically between $\tau = 0.001$ and 1000 fm/c and tuned the initial anisotropy to $\alpha_0 \simeq 0.0025$ following a method similar to the one outlined in the appendix of Ref.~\cite{Romatschke:2017vte}.  As can be seen from this figure, all solutions approach the attractor solution in a finite time. The slowest approach appears to be for moments with $m=0$ which appear in the leftmost column of Fig.~\ref{fig:attractorGridSols}.  Considering, for example, $\barM^{30}$, the generic solutions visibly merge with the attractor only after $\overline{w} \gtrsim 6$.  For $m \neq 0$, however, one sees that all moments computed from individual solutions visibly merge with the attractor after $\overline{w} \gtrsim 3$.   

I will make these statements more quantitative shortly, but before that I would like to discuss the source of the slow thermalization and approach to the attractor for the moments with $m=0$.  This subset of moments contains no power of $p_z$ in the integrand and so are sensitive to the behavior of the one-particle distribution in the vicinity of \mbox{$p_z=0$}.  In order to understand this behavior better, one can obtain the full one-particle distribution from the exact solution using Eq.~\eqref{eq:exactsolf}.  For this purpose, I first obtain the solution to the integral equation for the temperature \eqref{solT1} and then evaluate Eq.~\eqref{eq:exactsolf} in a grid in $p_{x,y}/T$ and $p_z/T$.  For the results show herein, I used a 500 $\times$ 500 grid in the scaled transverse and longitudinal momentum with $|p_{T,z}|/T \leq 50$.  

\begin{figure*}[t!]
\centerline{
\includegraphics[width=1\linewidth]{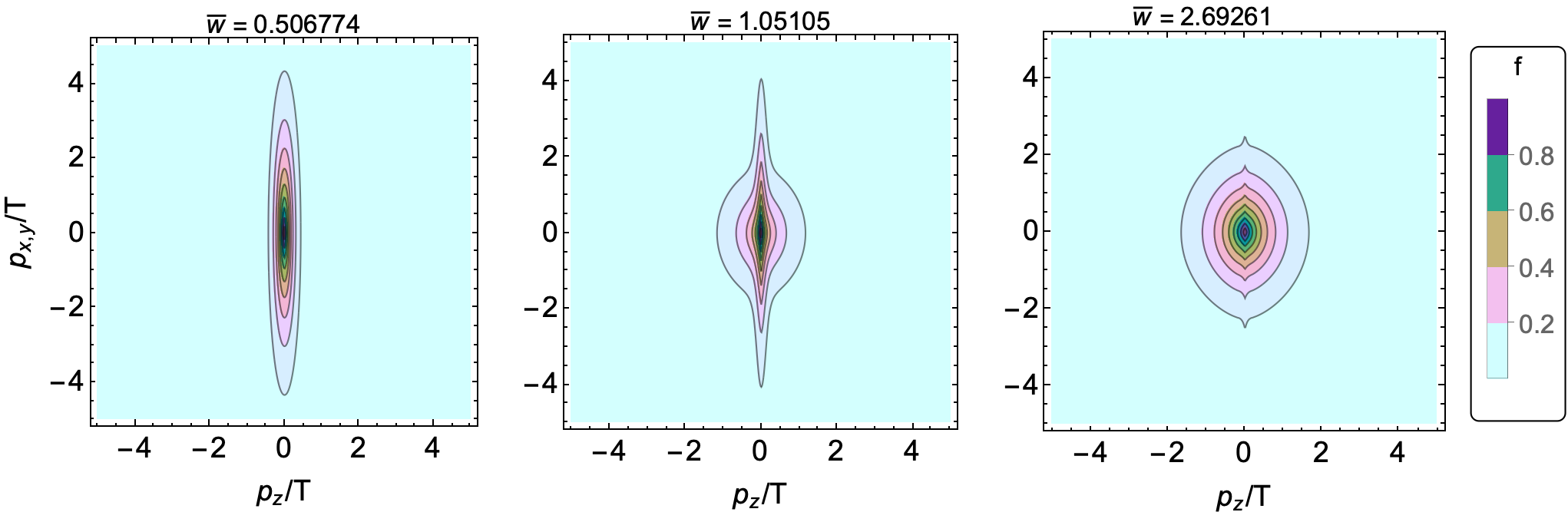}
}
\caption{Visualization of the one-particle distribution function obtained using a typical (non-attractor) anisotropic initial condition.}
\label{fig:f_vis1}
\end{figure*}

\begin{figure*}[t!]
\centerline{
\includegraphics[width=1\linewidth]{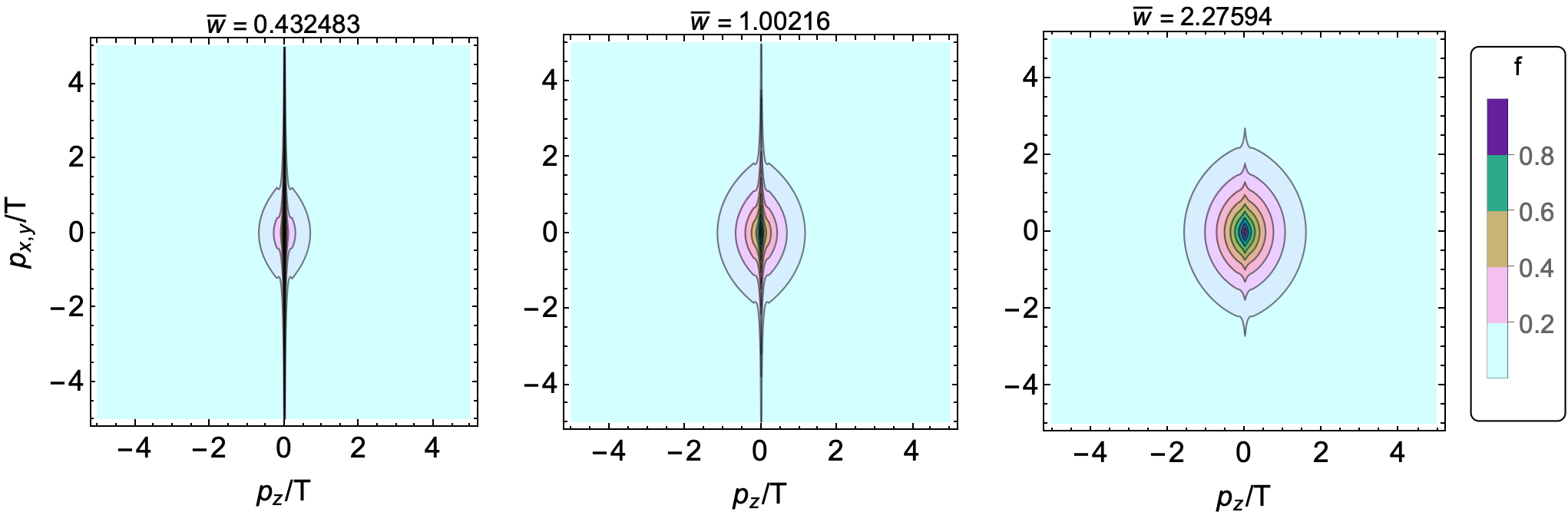}
}
\caption{Visualization of the one-particle distribution function associated with the attractor.}
\label{fig:f_vis2}
\end{figure*}
 
In Fig.~\ref{fig:f_vis1} I show snapshots of the one-particle distribution function at three different scaled times.  For this figure, I used a typical anisotropic initial condition from the set shown in Fig.~\ref{fig:attractorGridSols}.  As can be seen from this figure, the exact solution for the one-particle distribution function generically contains two visually identifiable components.  The first is an anisotropic piece which becomes increasingly more compressed into the region with $p_z \sim 0$ as a function of scaled time.  This contribution comes from the first term in the exact solution Eq.~\eqref{eq:exactsolf} which corresponds to the free streaming contribution.  As a function of time this contribution becomes more squeezed in the longitudinal direction, however, eventually the amplitude of this very narrow ridge decreases exponentially in time due to the damping function $D$ in the first term in Eq.~\eqref{eq:exactsolf}.  The second, more isotropic, component which can be seen in Fig.~\ref{fig:f_vis1} comes from the second term in \eqref{eq:exactsolf}.  This contribution dominates at late times.  Turning now to Fig.~\ref{fig:f_vis2}, in this figure I present similar snapshots of the one-particle distribution function, however, this time using the temperature evolution obtained from the attractor solution to Eq.~\eqref{solT1}.  In this case, one sees similar features to the generic solution presented previously, however, one sees that in this case the attractor initial condition used and the free streaming term result in a set highly-squeezed modes that have nearly $p_z=0$.  

\begin{figure*}[t!]
\centerline{
\includegraphics[width=0.5\linewidth]{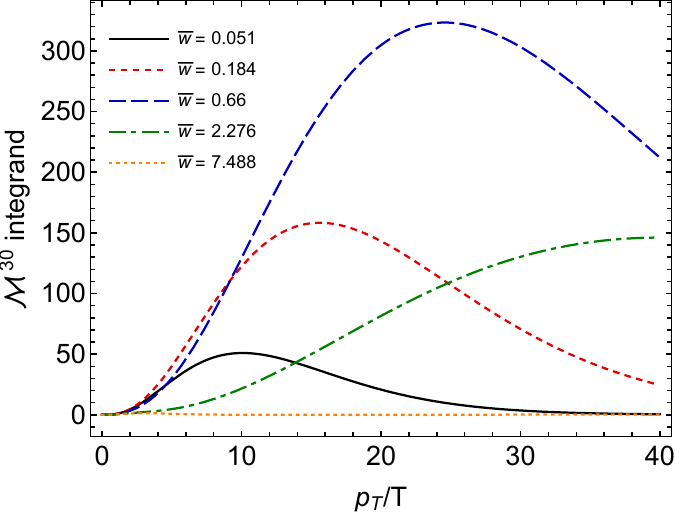}
}
\caption{Integrand of the ${\cal M}^{30}$ moment obtained from the exact attractor solution as a function of $p_T/T$ with $p_z=0$.  The different lines show different values of $\overline{w}$.}
\label{fig:m30plot}
\end{figure*}

It is precisely this set of squeezed modes which cause the approach to equilibrium to proceed more slowly for these moments than for moments with $m\neq0$.  For $m\neq0$, the powers of $p_z$ naturally reduce the impact of the highly squeezed modes.  For $m=0$ the moments will be dominated by these squeezed modes at early times and, as $n$ is increased, the magnitude of ${\cal M}^{n0}$ will increase for all $n > 2$ at early times.  As an example, in Fig.~\ref{fig:m30plot}, I plot the integrand of the ${\cal M}^{30}$ moment as a function of $p_T/T$ for $p_z=0$.  As this figure demonstrates, at early times, the integrand for this moment is dominated by modes with high transverse momentum.  The peak of the integrand moves to higher $p_T$ and has an increasing magnitude at early times, however, at late times, the amplitude of the peak is diminished due to the damping function which appears in the free streaming contribution to the one-particle distribution function.   Despite the presence of these highly squeezed modes, the dynamics of the longitudinal pressure and other low-order moments with $m>0$ are not significantly affected.

\begin{figure*}[t!]
\centerline{
\includegraphics[width=1\linewidth]{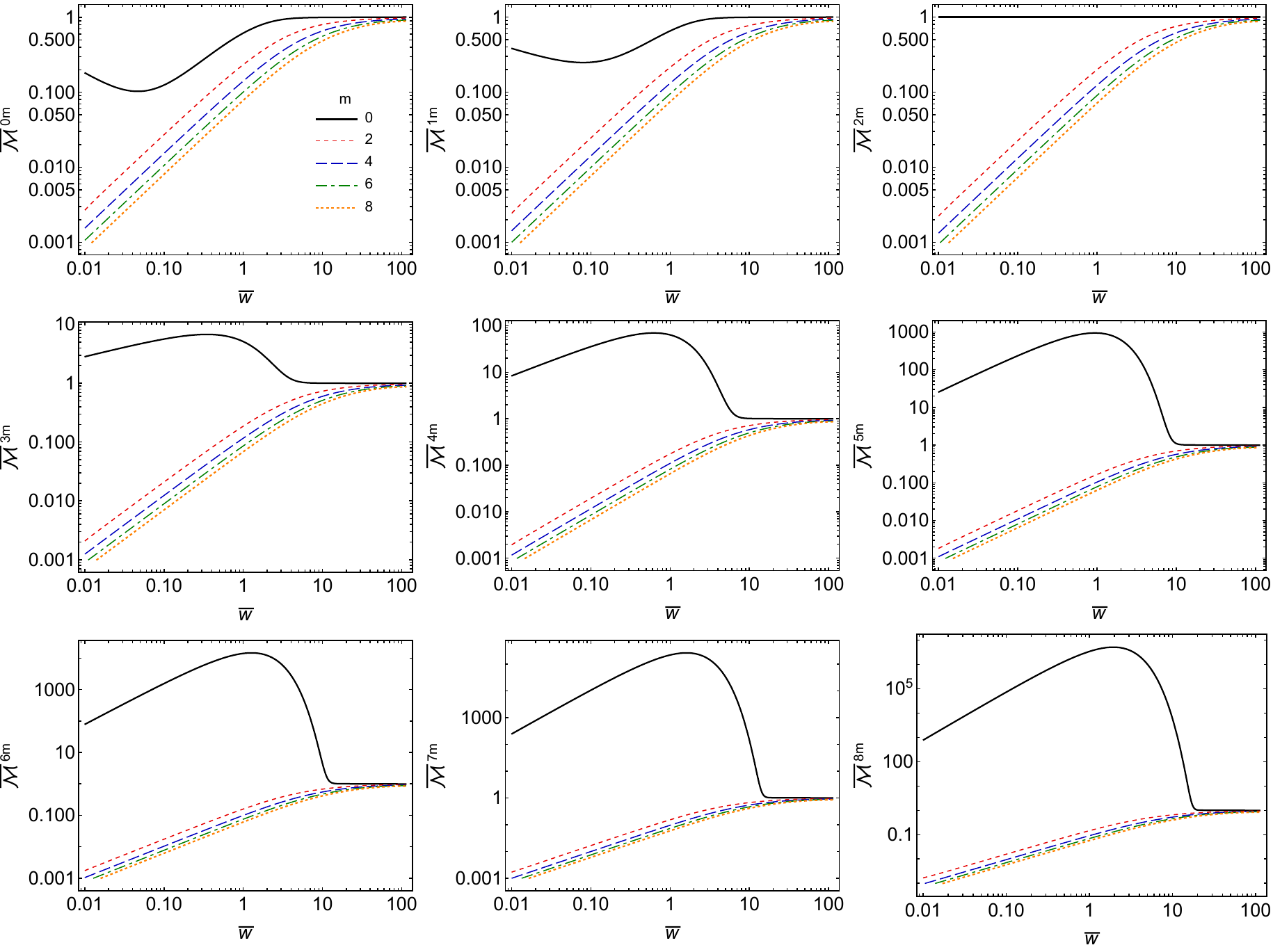}
}
\caption{Scaled moments $\barM^{nm}$ obtained from the exact attractor solution.  Panels show \mbox{$n \in \{0 \cdots 8\}$} and inside each panel the set of lines correspond to \mbox{$m \in \{0 \cdots 8\}$}.}
\label{fig:attractorGrid1}
\end{figure*}

\begin{figure*}[t!]
\centerline{
\includegraphics[width=1\linewidth]{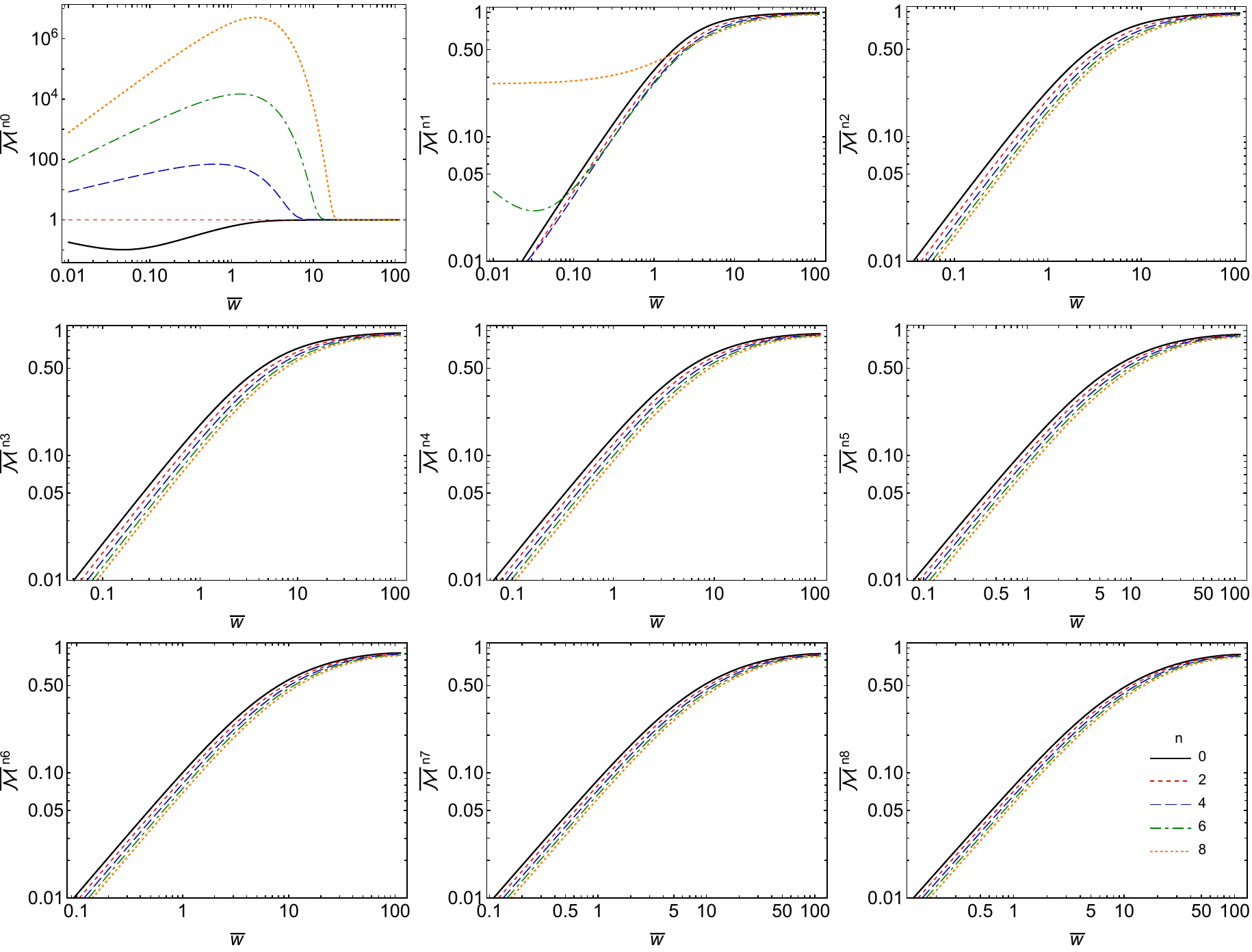}
}
\caption{Scaled moments $\barM^{nm}$ obtained from the exact attractor solution.  Panels show \mbox{$m \in \{0 \cdots 8\}$} and inside each panel the set of lines correspond to \mbox{$n \in \{0 \cdots 8\}$}.}
\label{fig:attractorGrid2}
\end{figure*}

Returning to the behavior of the general moments obtained from the exact solution, in Figs.~\ref{fig:attractorGrid1} and \ref{fig:attractorGrid2}, I present plots of the scaled-time dependence of $\barM^{nm}$ \eqref{mbardef} for $n,m \in \{1,\cdots,8\}$.  In Fig.~\ref{fig:attractorGrid1}, each panel corresponds to a different value of $n$ and the lines within each panel correspond to different values of $m$.  In order to make the plot more readable, I have only plotted the even values of $m$.  As one sees from Fig.~\ref{fig:attractorGrid1}, for fixed $n$ and $m > 0$, increasing $m$ results in a larger deviation from unity which indicates that higher longitudinal momenta take longer to equilibrate.  The same results are plotted in Fig.~\ref{fig:attractorGrid2}, however, for this figure each panel corresponds to a fixed value of $m$ and the lines in each panel correspond to different values of $n$.  Focusing first on the upper left panel of Fig.~\ref{fig:attractorGrid2}, one sees that the prediction that $\barM^{n0}$ should increase in magnitude at early times as $n$ is increased due to squeezed modes is borne out by the results.  The other panels demonstrate that, for fixed $m$, increasing $n$ results in slower thermalization.  Again this is consistent with the slower relaxation of high momentum modes.  

\begin{figure*}[t!]
\centerline{
\includegraphics[width=0.5\linewidth]{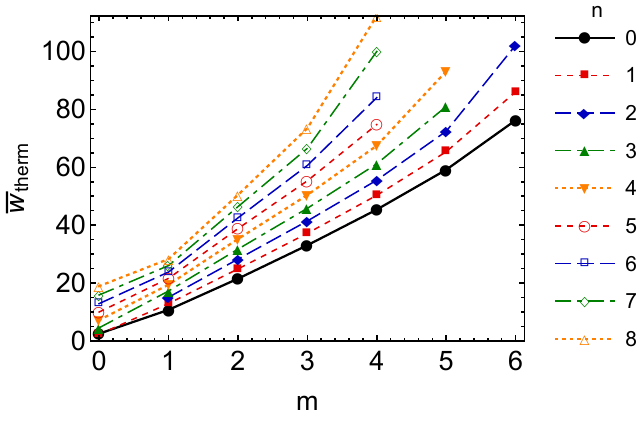}
}
\vspace{-3mm}
\caption{Scaled thermalization time $\overline{w}_{\rm therm}$ as function of $m$.  The different lines correspond to different values of $n$.}
\label{fig:thermalization1}
\end{figure*}

To make this more quantitative, one can extract a ``thermalization time'' $\overline{w}_{\rm therm}$ by solving for the scaled time at which a given moment falls within 10\% of its equilibrium value, i.e. when the $\barM^{nm} = 0.9$.  The results obtained are shown in Fig.~\ref{fig:thermalization1}. For this plot I exclude $\barM^{20}$, since this is always equal to unity by energy conservation.  In addition, if the thermalization time was longer that the maximum rescaled time (in this case $\overline{w}_{\rm max} \simeq 112$), then I don't plot a value.  As Fig.~\ref{fig:thermalization1} demonstrates, the thermalization time increases as both $n$ and $m$ are increased.  Again, this is indicative of the slower thermalization of high-momentum modes since the integrands for large $n$ and $m$ are dominated by large momenta.  I will present direct evidence for this conclusion later using the exact solution for the one-particle distribution function.

\begin{figure*}[t!]
\centerline{
\includegraphics[width=1\linewidth]{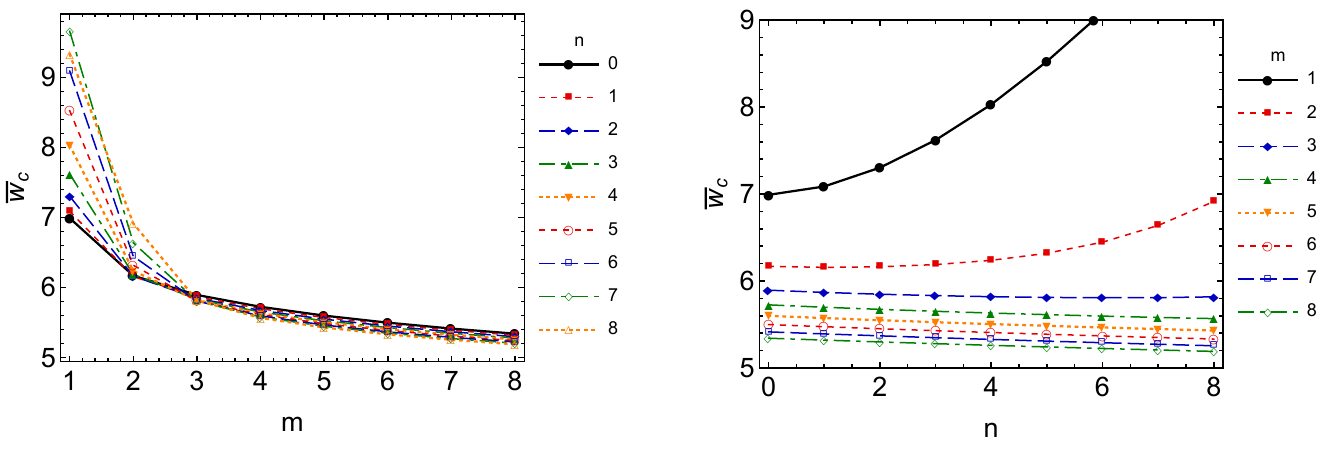}
}
\vspace{4mm}
\centerline{
\includegraphics[width=0.4\linewidth]{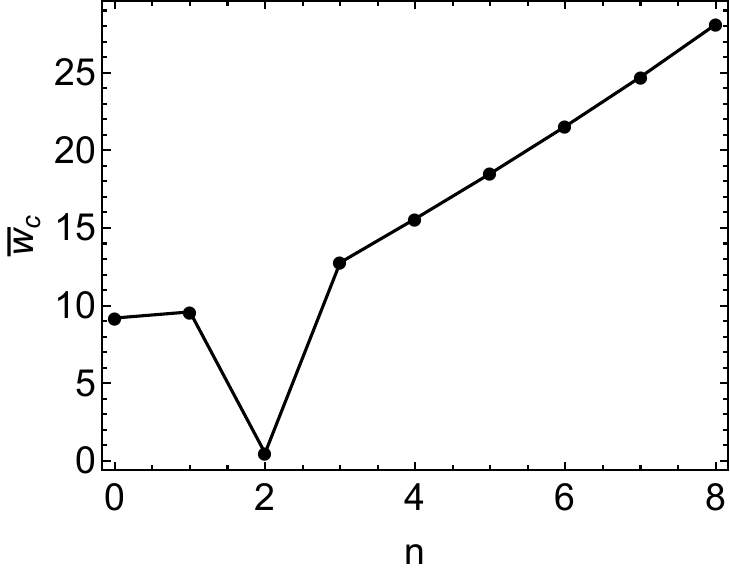}
}
\caption{The pseudo-thermalization time $\overline{w}_c$ for $n,m \in \{1,\cdots,8\}$.  The top left panel shows $\overline{w}_c$ as a function of $m$ with the lines corresponding to different values of $n$.  The top right panel shows $\overline{w}_c$ as a function of $n$ with the lines corresponding to different values of $m$.  The bottom panel shows $\overline{w}_c$ as a function of $n$ for the case $m=0$. }
\label{fig:hydrodynamization}
\end{figure*}

Another question that naturally arises is what is the time scale for a moment $\barM^{nm}$ obtained from a generic initial condition to approach the corresponding exact attractor solution.  For this purpose, I computed the difference between the individual runs and attractor solution $\barM^{nm}_{\rm attractor}$ shown in Fig.~\ref{fig:attractorGridSols}.  I then solved for the time at which all solutions converged to the attractor solution to within $10^{-6}$, i.e. I require $\max|\barM^{nm}_{\rm i}(\overline{w}_{c}) - \barM^{nm}_{\rm attractor}(\overline{w}_{c}) | < 10^{-6}$ where $i$ indexes each of the individual runs.  The resulting ``pseudo-thermalization time" is called $\overline{w}_c$ and is plotted in Fig.~\ref{fig:hydrodynamization}.  In Fig.~\ref{fig:hydrodynamization}, the top left panel shows $\overline{w}_c$ as a function of $m$ with the set of lines corresponding to different values of $n$.  As can be seen from this figure, all moments with $m\geq1$ have short pseudo-thermalization times that decrease with increasing $m$ in the range shown.  This is surprising since in Fig.~\ref{fig:thermalization1} one sees that increasing $m$ for fixed $n$ results in larger thermalization times.  Turning to the top right panel of Fig.~\ref{fig:hydrodynamization}, one sees once again that, for the moments shown, one has $\overline{w}_c < \overline{w}_{\rm therm}$, with their separation increasing as one increases $m$ and $n$ for $m > 2$.  The moments with $m=0,1,2$ are found to have pseudo-thermalization times that increase with increasing $n$.  Because of the limited range shown, it is possible that other moments might start increasing at very large $n$.  Finally, because of their slow relaxation, in the bottom panel of Fig.~\ref{fig:hydrodynamization} I plot $\overline{w}_c$ as a function of $n$ for moments with $m=0$, separately.\footnote{For $\barM^{20}$, I chose $\overline{w}_c$ to be the minimum time since $\barM^{20}=1$ at all times by Landau matching.}  From this panel one sees evidence that, for $m=0$, the pseudo-thermalization time increases approximately linearly large $n$.  Comparing with Fig.~\ref{fig:thermalization1} one sees that, for the $m=0$ moments, it is possible to have $\overline{w}_c \sim \overline{w}_{\rm therm}$ or to even invert their order.

\begin{figure*}[t!]
\centerline{
\includegraphics[width=0.97\linewidth]{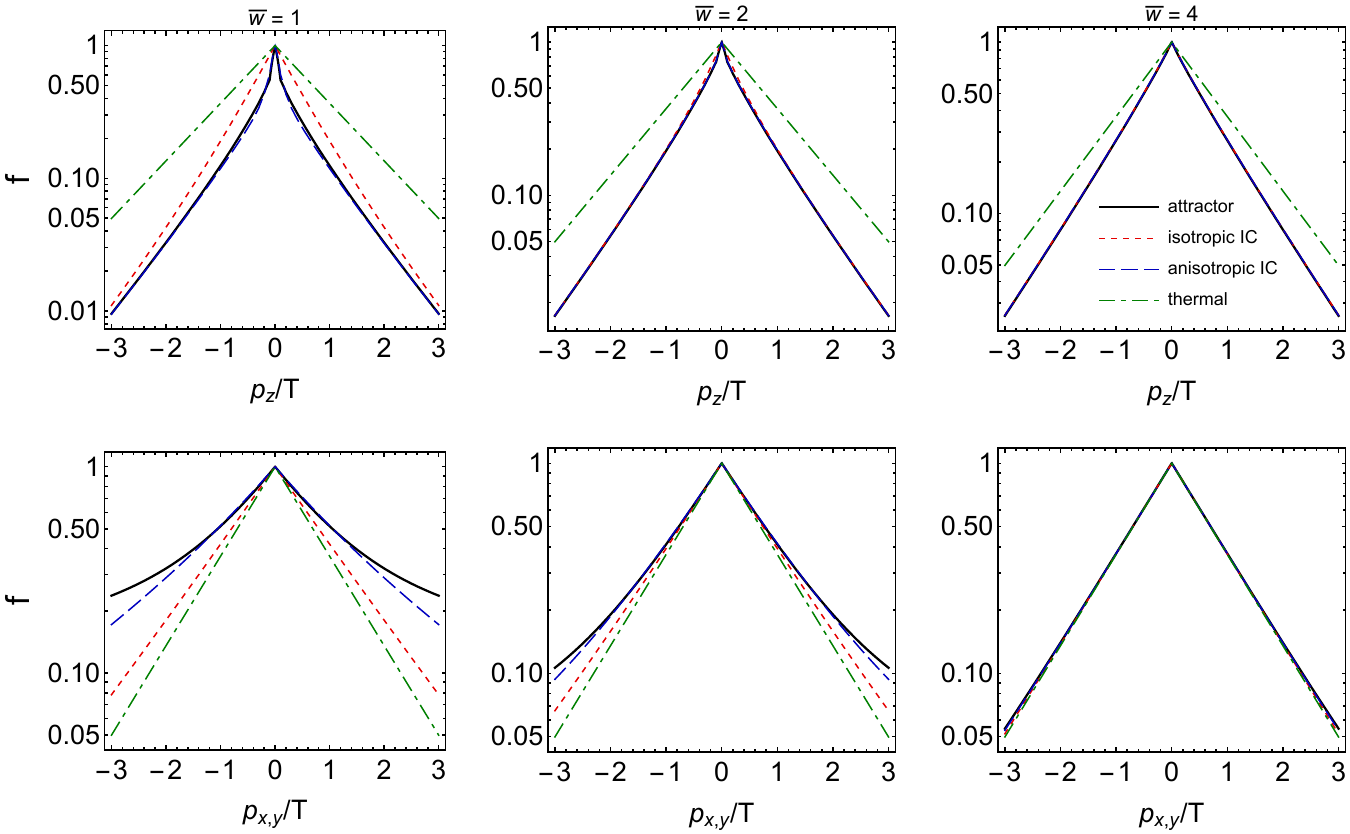}
}
\caption{Attractor for the full distribution function.  The solid black line is the attractor result, the red short-dashed line is a typical exact solution with isotropic initial conditions, the blue long-dashed line is a typical exact solution with anisotropic initial conditions, and the green dot-dashed is a thermal distribution using the effective temperature obtained from the exact attractor solution.  The top row shows $f(p_T,p_z=0)$ and the bottom row shows $f(p_T=0,p_z)$.  The columns correspond to different scaled times $\overline{w}$.}
\label{fig:fAttractor1}
\end{figure*}
 
\begin{figure*}[t!]
\centerline{
\includegraphics[width=0.97\linewidth]{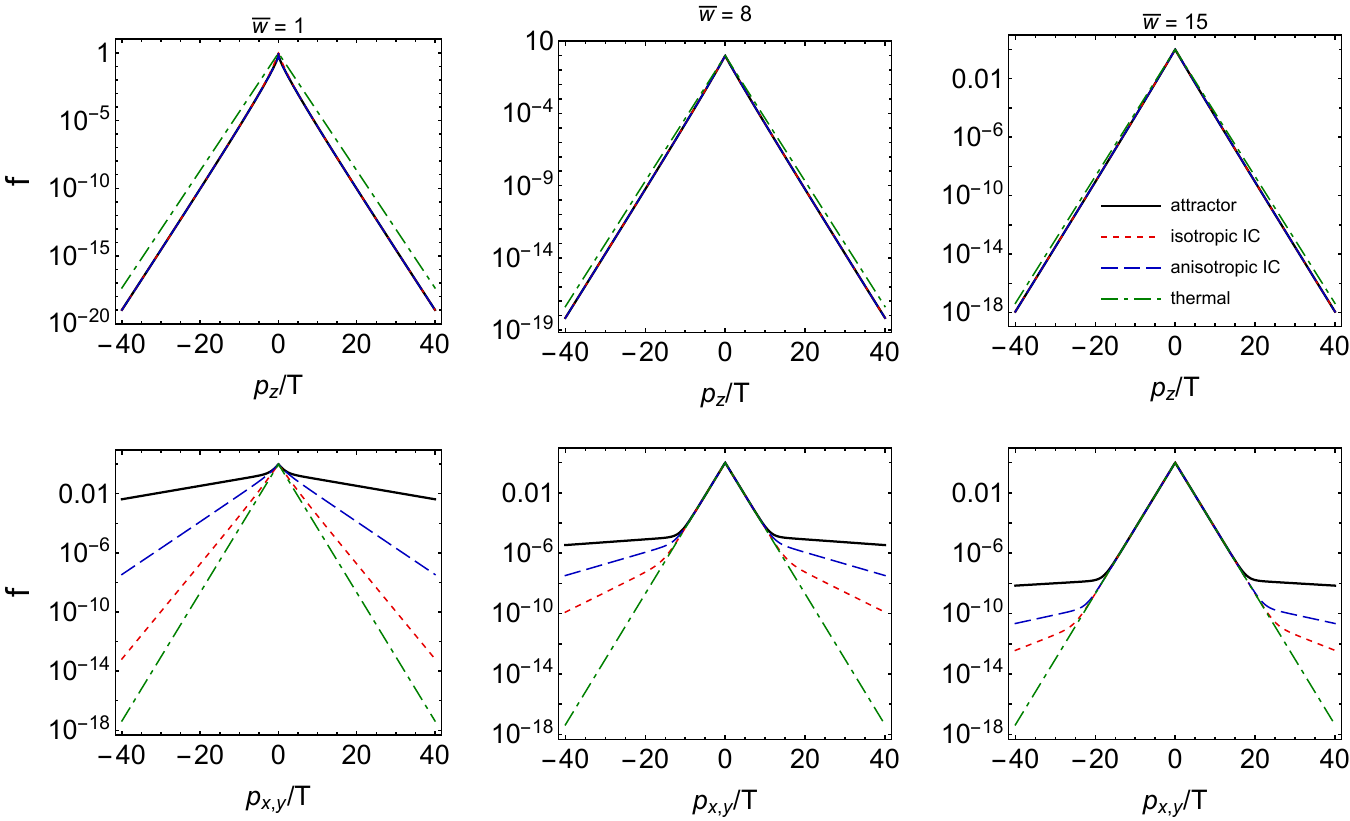}
}
\caption{Attractor for the full distribution function at larger momenta.  The labeling etc. is the same as in Fig.~\ref{fig:fAttractor1}.}
\label{fig:fAttractor2}
\end{figure*}

Summarizing, one sees that in the range $0 \leq n \leq 8$, $1 \leq m \leq 8$ one has $\overline{w}_c < \overline{w}_{\rm therm}$ with the separation increasing with $m$ and $n$.  From the results, one also sees indications that for $1 \leq m \leq 2$, $\overline{w}_c$ increases with $n$ for large $n$, but slower than $\overline{w}_{\rm therm}$ increases.  For moments with $m=0$, one sees large $\overline{w}_c$ which can become on the order of or exceed $\overline{w}_{\rm therm}$.  Based on this, it's hard to draw firm conclusions about the nature of the (psuedo)-thermalization of the system for all possible moments.  To shed some light on this question, one can compute the full distribution function for the exact solution using the attractor initial conditions.  As mentioned previously, once the solution to the integral equation for the temperature \eqref{solT1} is obtained, one can evaluate Eq.~\eqref{eq:exactsolf} in a grid in $p_{x,y}/T$ and $p_z/T$.

The result for the distribution function associated with the attractor is visualized in Fig.~\ref{fig:f_vis2}, however, it's hard to draw quantitative conclusions from such plots.  In Figs.~\ref{fig:fAttractor1} and \ref{fig:fAttractor2}, I present the distribution function along the line with $p_T=0$ as a function of $p_z$ (top row) and along the line with $p_z=0$ as a function of $p_T$ (bottom row).  Fig.~\ref{fig:fAttractor1} shows the low-momentum region $p_{x,y,z}/T \leq 3$ and Fig.~\ref{fig:fAttractor1} shows the high-momentum region $p_{x,y,z}/T \leq 40$.  Focusing on Fig.~\ref{fig:fAttractor1}, firstly one sees that the two typical solutions (red short-dashed and blue long-dashed lines, respectively) converge towards the attractor solution (black solid line), however, the approach is not uniform in the sense that in some regions of momentum the solutions approach the attractor from below while in other regions they approach it from above.  Secondly, one sees that the approach to attractor and thermal solutions (black solid and green dot-dashed lines, respectively) is slower along the transverse line ($p_z=0$) than along the longitudinal line ($p_{x,y}=0$).  This is consistent with there being a subset high momentum modes with high $p_T$ and nearly $p_z=0$ which equilibrate more slowly than the rest of the modes.  

\begin{figure*}[t!]
\centerline{
\includegraphics[width=0.95\linewidth]{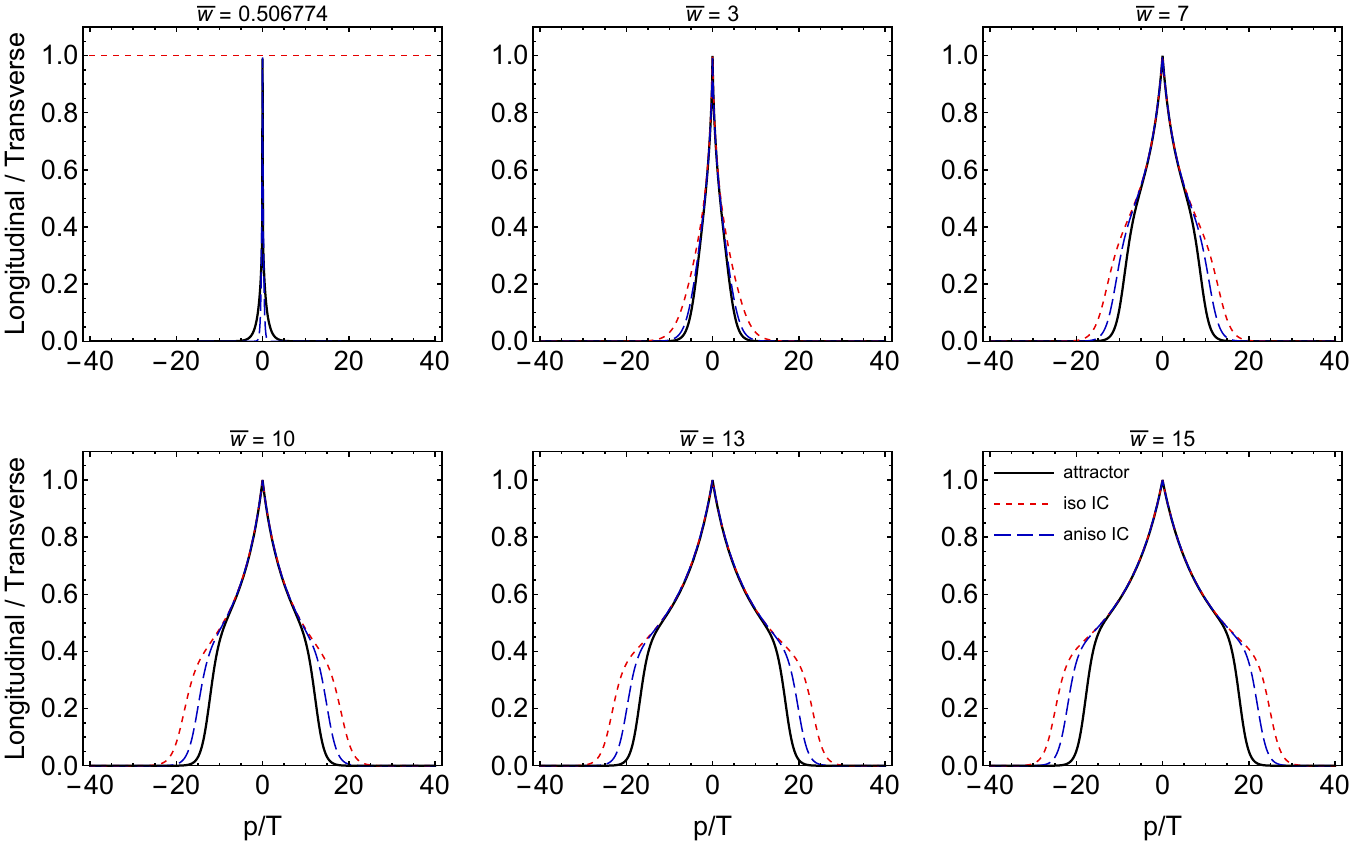}
}
\caption{Longitudinal to transverse ratio, $f(0,0,p)/f(p,0,0)$, as a function of $p/T$.  The different panels correspond to different scaled times $\overline{w}$ and within each panel the ratio obtained from the attractor is shown as a solid line while the ratio obtained from typical isotropic and anisotropic initial conditions are plotted as red short-dashed and blue long-dashed lines, respectively.}
\label{fig:plotsLT}
\end{figure*}

This can be seen more clearly if one zooms out and investigates the behavior of the distribution function at higher momentum, as shown in Fig.~\ref{fig:fAttractor2}.  From this figure, one sees that indeed the distribution in the longitudinal direction (top row) approaches the attractor solution and thermal approximation very quickly, whereas along the transverse direction (bottom row) one sees a two-component distribution function, with a subset of low-momentum modes which quickly approach the attractor being visually distinct from modes with high-momentum that have a much more shallow exponential slope.  In addition, one sees a sharp transition from the low-momentum to the high-momentum behavior and that this transition point, $p_{\rm transition}$, moves to higher momentum as a function of $\overline{w}$.

To investigate the behavior of the transition point $p_{\rm transition}$  more quantitatively one can take the ratio of the distribution function evaluated along the line with $p_{x,y}  = 0$ (longitudinal direction) to the same along the line with $p_z=0$ (transverse direction), i.e. $f(0,0,p)/f(p,0,0)$. I present plots of this ratio in Fig.~\ref{fig:plotsLT}.  As can be see from this figure, there is an isotropization ``front'' that moves to the right and which can be associated with the position of the transition point $p_{\rm transition}$  in the distribution function visible in the bottom row of Fig.~\ref{fig:fAttractor2}.  If one defines the transition point as the point at which the longitudinal to transverse ratio along the transverse directions equals 1/2, one finds that, for the attractor solution, at $\overline{w} \gtrsim 2$ the hydrodynamization front moves out in temperature-scaled momentum at a constant speed of 1.5, i.e. $p_{\rm transition}/T \simeq 1.5 \, \overline{w}$.  A similar hydrodynamization front can be seen in the typical solutions (dashed lines), however, in this case one finds that the hydrodynamization front speed is faster than 1.5 and approaches 1.5 at asymptotically late times as a given typical solution converges to the attractor solution.

\begin{figure*}[t!]
\centerline{
\includegraphics[width=1\linewidth]{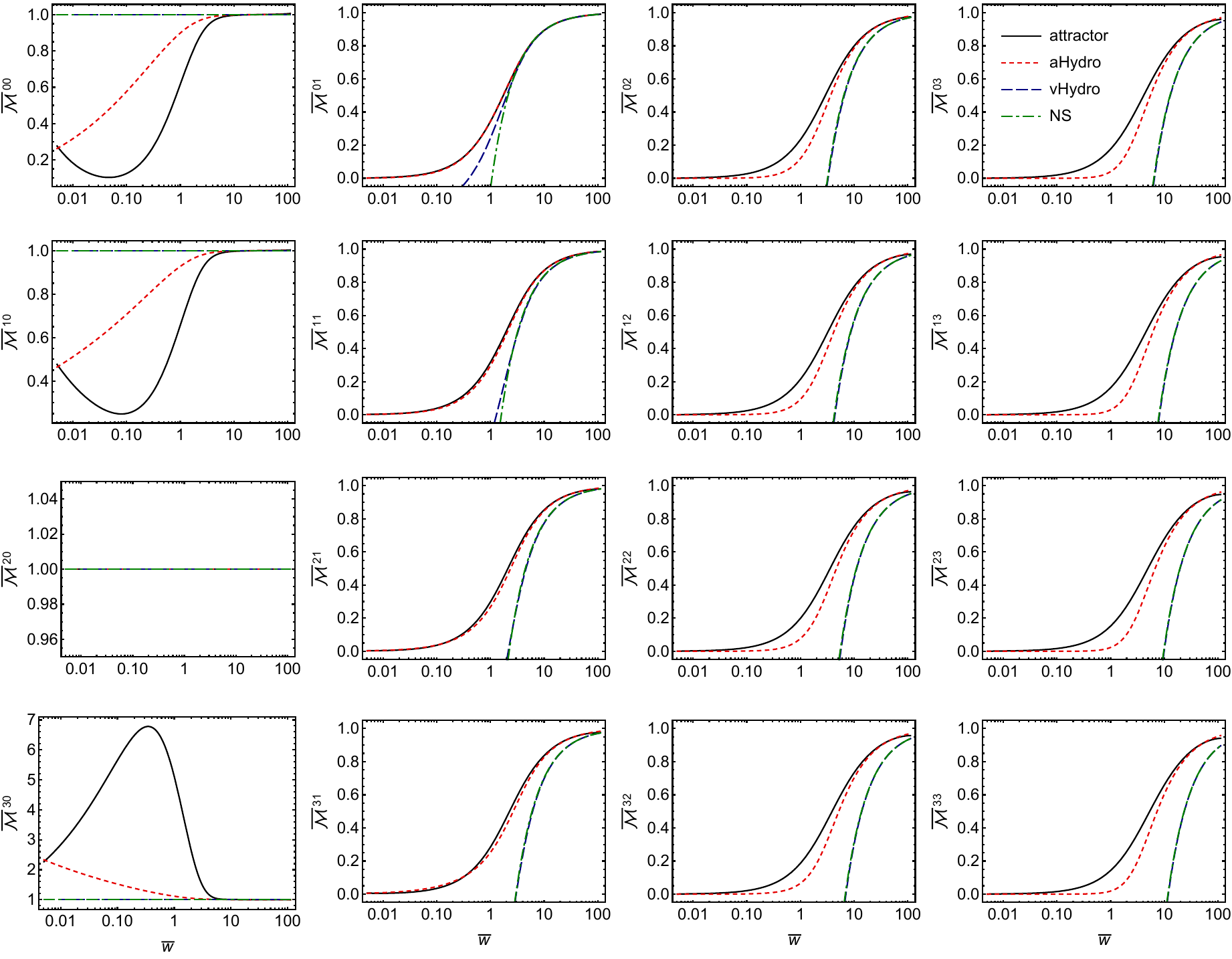}
}
\caption{Scaled moments $\barM^{nm}$ obtained from the exact attractor solution (solid black line) compared with the aHydro attractor (red dashed lines), DNMR attractor (blue long dashed lines), and the Navier-Stokes limit for each moment (green dot-dashed lines).  Horizontal axis is $\bar{w} \equiv \tau T/5 \bar\eta$.  Panels show a grid in $n$ and $m$.}
\label{fig:attractorGridComp}
\end{figure*}

Finally, I would like to compare the exact results obtained for a subset of moments with results obtained from the aHydro and vHydro dissipative hydrodynamics approximations.  For vHydro, I determine the attractor using the complete second-order viscous hydrodynamics equations of Denicol, Niemi, Molnar, and Rischke (DNMR) \cite{Denicol:2010xn,Denicol:2011fa}.  For aHydro, I use the method introduced originally by Florkowski and Tinti \cite{Tinti:2013vba} which utilizes the first and second moments of the Boltzmann equation.  For details concerning the determination of the attractor for both aHydro and vHydro, I refer to the reader to Ref.~\cite{Strickland:2017kux}.  In both cases, the attractor is determined from the solution of a one-dimensional ordinary one-dimensional ordinary differential equation subject to the appropriate initial condition.  For vHydro, one extracts $\bar\pi = \pi/\epsilon$ and, using this, one can reconstruct the solution for any moment using Eq.~\eqref{eq:vhydromoms}.  From the vHydro result, one can also obtain the Navier-Stokes (NS) result by taking $\bar\pi = 16 \bar\eta/(9 \tau T)$.  For aHydro, one extracts $\alpha$ associated with the attractor solution and uses Eq.~\eqref{eq:ahydromoms} to compute the moments.  In Fig.~\ref{fig:attractorGridComp}, I compare the exact attractor (black solid lines) with the aHydro attractor (red dashed lines), DNMR attractor (blue long dashed lines), and the NS limit for each moment (green dot-dashed lines).  In all cases shown, aHydro provides a better approximation to the exact moments than the vHydro or NS solutions.  

In the case of aHydro, one sees that the worst agreement is for the $m=0$ moments (leftmost column of Fig.~\ref{fig:attractorGridComp}).  This can be understood from the fact that within aHydro one assumes that the one-particle distribution function is given by a single spheroid in momentum space.  As a result, aHydro fails to accurately describe the evolution of the modes with $m=0$ which are dominated by the free-streaming part of the evolution and remain highly anisotropic at all times.  For $m>0$ one sees that, as $m$ and $n$ are increased, the aHydro results differ more and more from the exact solutions in the transition region around $\overline{w} \sim 1-10$, however, the aHydro solution does not appear to ``break'' in the sense of giving unphysical (e.g. negative) results for any of the moments considered.\footnote{Eq.~\eqref{eq:ahydromoms} guarantees the positivity of all moments by construction.}

\begin{figure*}[t!]
\centerline{
\includegraphics[width=0.95\linewidth]{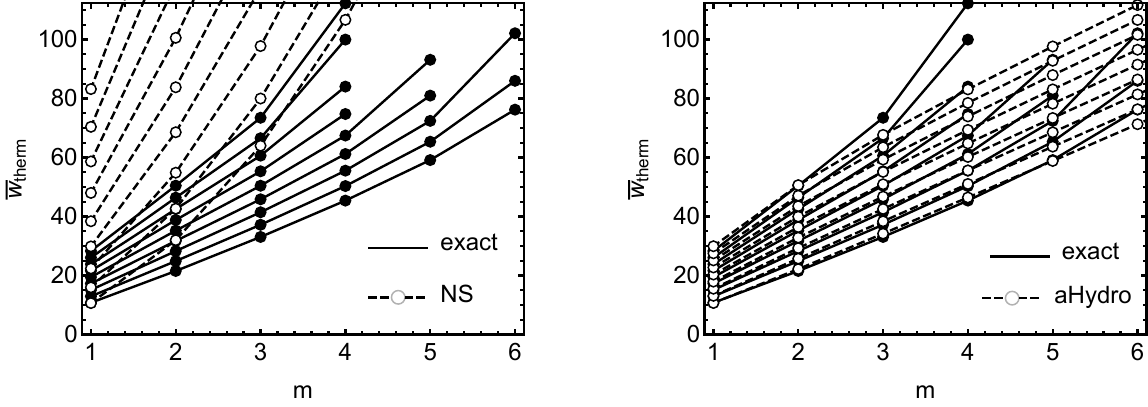}
}
\caption{Scaled thermalization time $\overline{w}_{\rm therm}$ as function of $m$.  The different lines correspond to different values of $n$.  The left panel shows a comparison with vHydro and the right panel shows a comparison with aHydro.}
\label{fig:thermalization2}
\end{figure*}

Turning to the vHydro and NS results, firstly one sees that for $m=0$ both schemes predict $\barM^{n0}=1$ at all times.  This is by construction in vHydro since the $\delta f$ correction to the vHydro distribution function vanishes in order to guarantee that the number density and energy density are unaffected by the viscous correction.  As a consequence, vHydro and NS do not provide reliable approximations for these moments for $n \neq 2$.  For $m>0$, one sees that, although the vHydro and NS results do a reasonable job in describing the low $n$ and $m$ moments, as $n$ and $m$ are increased these approximations become significantly worse.  In fact, one sees that for all moments with $m>0$, both vHydro and NS predict that the various moments become negative at early times.  This is due to the breakdown of the near-equilibrium assumption and violates positivity of the moments from Eq.~\eqref{eq:genmom1}.  The problem becomes more severe as one increases $m$ and $n$ due to the increasing powers of $p_z$ in the moment integrands.  Relatedly, in Fig.~\ref{fig:thermalization2}, I compare the thermalization time $\overline{w}_{\rm therm}$ obtained from the exact solution (see Fig.~\ref{fig:thermalization1} for the values of $n$ for each curve) with the NS solution (left panel) and the aHydro solution (right panel).  As can be seen from this figure, as $m$ and $n$ increase standard viscous hydrodynamics diverges from the exact solution, whereas aHydro continues to provide a reasonable approximation for the thermalization time determined from the exact solution for each moment.

\section{Conclusions and outlook}
\label{sect:conclusions}

In this paper, I have extended the analysis of the exact attractor for the 0+1d RTA Boltzmann equation to include higher-order moments of the one-particle distribution function and the one-particle distribution function itself.  Starting with the latter, Figs.~\ref{fig:fAttractor1} and \ref{fig:fAttractor2} demonstrate that the one-particle distribution exhibits attractor-like behavior in that generic solutions converge to an attractor on a characteristic time scale, which I have dubbed the ``pseudo-thermalization time'', $\overline{w}_c$.  I demonstrated that, within RTA, the pseudo-thermalization time depends primarily on the transverse momentum cut considered and that the longitudinal momenta converge to their attractor on a much faster time scale.  The fact that the transverse modes isotropize more slowly was shown to originate from the free streaming contribution to the exact solution, which contributes a narrow (highly momentum-space anisotropic) band of modes in the vicinity of $p_z \sim 0$ which dominate moments of the one-particle distribution function which do not contain powers of $p_z$, i.e. $\barM^{n0}$.  Contrarily, I demonstrated that moments with $m>0$ have pseudo-thermalization times that are parametrically shorter than the corresponding thermalization time for $1 \leq m \leq 8$ and $0 \leq n \leq 8$.  I presented a detailed comparison of the moments in this range and provided numerical estimates for both the thermalization and pseudo-thermalization times as a function of $m$ and $n$.

Due to the fact that evolution of a large set of moments is sometimes difficult to interpret, I also presented solutions for the distribution function obtained from the exact RTA attractor solutions.  Using these solutions, I demonstrated that there is an hydrodynamization front that propagates out along the $p_z=0$ plane from low-momenta to high-momenta with a constant speed,  $p_T/T \simeq 1.5 \, \overline{w}$.  With this in hand, the behavior of the various moments becomes easier to understand, in particular, the behavior of the $\barM^{n0}$ moments.  Based on these findings, one can determine the time scale for momenta up to $p_T \sim 2$ GeV to thermalize.  Assuming a hadronic freeze-out temperature of $T_{\rm FO} = 150$ MeV, then requires that momenta $p_T/T \lesssim 13$ are approximately thermalized.  Using the propagation speed for the hydrodynamization front then gives $\overline{w} \gtrsim 9$.  To convert this to a physical time, one can use the exact solution with the following (typical LHC) initial conditions $\tau_0 = 0.25$ fm/c, \mbox{$T_0 = 0.6$ GeV}, and $\eta/s = 0.2$, giving \mbox{$\tau \gtrsim 8$ fm/c}.  One can compare this to the time it takes the system to reach the freeze-out temperature, which using the same initial conditions, is \mbox{$\tau_{\rm FO}  \simeq$ 23 fm/c}.  Relatedly, one can determine the maximum transverse momentum which is thermalized at freeze-out. Using the same initial conditions, one finds $\overline{w}_{\rm FO} \simeq  17.3$ giving $p_T \lesssim 3.9$ GeV.

Finally, I presented a comparison of the attractors for moments of the one-particle distribution function using the aHydro, DNMR second-order vHydro, and Navier-Stokes approximations to the exact attractor solution.  For all three hydrodynamic schemes, I was able to find compact analytical expressions that could be used to compute any moment based on the exact solution for the pressure anisotropy (or equivalently the amplitude $\varphi$) in the corresponding scheme.  I found that the aHydro attractor for a general moment was the best approximation to exact attractor and that it also provided reliable estimates for the thermalization time of higher-order moments.  One obvious shortcoming of aHydro is evident from this work, namely the use of a single spheroid.  The exact solution for the distribution function has a two-component form with a highly-anisotropic free-streaming contribution and isotropizing component.  It would be interesting to see if one could generalize aHydro to include such a two-component distribution.  Turning, in the end, to first- and second-order viscous hydrodynamics, I found that, while they provide good approximations for low-order moments, higher-order moments are poorly reproduced and the resulting estimates for the pseudo-thermalization time for the higher-moments do not agree with the exact solutions.  For this reason, when considering higher moments, one can no longer associate the pseudo-thermalization time with the ``hydrodynamization time''.

Looking to the future, it would be interesting to understand more about the numerically observed hydrodynamization front.  I presented evidence that the propagation speed for the attractor solution is 3/2 in dimensionless units.  It would be nice to have an analytical understanding of this result.  Along these lines, Heller, Kurkela and Spalinski \cite{Heller:2016rtz} demonstrated that the solutions to the exact RTA kinetic integral equation \eqref{solT1} possess infinitely many transient modes that carry the majority of information about the distribution function entering the energy density at late times.  Since the evolution of the distribution function \eqref{eq:exactsolf} can be written entirely in terms of the solution to  \eqref{solT1}, the modes identified by Heller et al should also govern the evolution of the full distribution function and, hence, all moments.  This seems to provide a natural explanation of the emergence of an attractor for all moments.  It would be interesting to study the evolution of higher moments and the distribution function itself using their approach.  Other natural extensions of this work include, for example, studying the dependence on the underlying statistics of the isotropic distribution function similar to Ref.~\cite{Florkowski:2014sda,Florkowski:2015lra} and studying the effect of enforcing number conservation for a single component scalar field or a multi-component quark-gluon system, as done in Ref.~\cite{Florkowski:2017jnz}.
 
\section*{Corrections}

In this version (arXiv v3), I have corrected an error which affected Figs.~\ref{fig:f_vis1}, \ref{fig:f_vis2}, \ref{fig:fAttractor1}, \ref{fig:fAttractor2}, and \ref{fig:plotsLT}.  This stemmed from a bug in the original reconstruction of the full distribution function. All other figures remain unchanged.  I thank the authors of Ref.~\cite{Jaiswal:2021uvv}, in particular C. Chattopadhyay, for calling this to my attention.  Luckily, all discussions and conclusions are unaffected by the change in the content of these figures.  For accuracy of the description, I have taken this opportunity to change the nomenclature "isotropization front" to "hydrodynamization front".

\acknowledgments

I thank J. Casalderry-Solana and U. Tantary for discussions.  I was supported by the U.S. Department of Energy, Office of Science, Office of Nuclear Physics under Award No. DE-SC0013470.

\bibliographystyle{JHEP}
\bibliography{attractor2}

\providecommand{\href}[2]{#2}\begingroup\raggedright\begin{thebibliography}{10}

\bibitem{Baier:2000sb}
R.~Baier, A.~H. Mueller, D.~Schiff and D.~Son, \emph{{'Bottom up'
  thermalization in heavy ion collisions}},
  \href{https://doi.org/10.1016/S0370-2693(01)00191-5}{\emph{Phys.Lett.}
  {\bfseries B502} (2001) 51}
  [\href{https://arxiv.org/abs/hep-ph/0009237}{{\ttfamily hep-ph/0009237}}].

\bibitem{Blaizot:2001nr}
J.-P. Blaizot and E.~Iancu, \emph{{The Quark gluon plasma: Collective dynamics
  and hard thermal loops}}, {\emph{Phys.Rept.} {\bfseries 359} (2002) 355}
  [\href{https://arxiv.org/abs/hep-ph/0101103}{{\ttfamily hep-ph/0101103}}].

\bibitem{Romatschke:2003ms}
P.~Romatschke and M.~Strickland, \emph{Collective modes of an anisotropic quark
  gluon plasma}, {\emph{Phys. Rev.} {\bfseries D68} (2003) 036004}
  [\href{https://arxiv.org/abs/hep-ph/0304092}{{\ttfamily hep-ph/0304092}}].

\bibitem{Arnold:2003rq}
P.~B. Arnold, J.~Lenaghan and G.~D. Moore, \emph{{QCD plasma instabilities and
  bottom up thermalization}}, {\emph{JHEP} {\bfseries 0308} (2003) 002}
  [\href{https://arxiv.org/abs/hep-ph/0307325}{{\ttfamily hep-ph/0307325}}].

\bibitem{Arnold:2004ti}
P.~Arnold, J.~Lenaghan, G.~D. Moore and L.~G. Yaffe, \emph{Apparent
  thermalization due to plasma instabilities in quark gluon plasma},
  {\emph{Phys. Rev. Lett.} {\bfseries 94} (2005) 072302}
  [\href{https://arxiv.org/abs/nucl-th/0409068}{{\ttfamily nucl-th/0409068}}].

\bibitem{Mrowczynski:2004kv}
S.~Mrowczynski, A.~Rebhan and M.~Strickland, \emph{{Hard loop effective action
  for anisotropic plasmas}},
  \href{https://doi.org/10.1103/PhysRevD.70.025004}{\emph{Phys.Rev.} {\bfseries
  D70} (2004) 025004} [\href{https://arxiv.org/abs/hep-ph/0403256}{{\ttfamily
  hep-ph/0403256}}].

\bibitem{Rebhan:2004ur}
A.~Rebhan, P.~Romatschke and M.~Strickland, \emph{{Hard-loop dynamics of
  non-Abelian plasma instabilities}},
  \href{https://doi.org/10.1103/PhysRevLett.94.102303}{\emph{Phys.Rev.Lett.}
  {\bfseries 94} (2005) 102303}
  [\href{https://arxiv.org/abs/hep-ph/0412016}{{\ttfamily hep-ph/0412016}}].

\bibitem{Rebhan:2005re}
A.~Rebhan, P.~Romatschke and M.~Strickland, \emph{Dynamics of quark-gluon
  plasma instabilities in discretized hard-loop approximation}, {\emph{JHEP}
  {\bfseries 09} (2005) 041}
  [\href{https://arxiv.org/abs/hep-ph/0505261}{{\ttfamily hep-ph/0505261}}].

\bibitem{Romatschke:2005pm}
P.~Romatschke and R.~Venugopalan, \emph{{Collective non-Abelian instabilities
  in a melting color glass condensate}},
  \href{https://doi.org/10.1103/PhysRevLett.96.062302}{\emph{Phys.Rev.Lett.}
  {\bfseries 96} (2006) 062302}
  [\href{https://arxiv.org/abs/hep-ph/0510121}{{\ttfamily hep-ph/0510121}}].

\bibitem{Romatschke:2006nk}
P.~Romatschke and R.~Venugopalan, \emph{{The Unstable Glasma}},
  \href{https://doi.org/10.1103/PhysRevD.74.045011}{\emph{Phys.Rev.} {\bfseries
  D74} (2006) 045011} [\href{https://arxiv.org/abs/hep-ph/0605045}{{\ttfamily
  hep-ph/0605045}}].

\bibitem{Romatschke:2006wg}
P.~Romatschke and A.~Rebhan, \emph{{Plasma Instabilities in an Anisotropically
  Expanding Geometry}},
  \href{https://doi.org/10.1103/PhysRevLett.97.252301}{\emph{Phys. Rev. Lett.}
  {\bfseries 97} (2006) 252301}
  [\href{https://arxiv.org/abs/hep-ph/0605064}{{\ttfamily hep-ph/0605064}}].

\bibitem{Rebhan:2008uj}
A.~Rebhan, M.~Strickland and M.~Attems, \emph{{Instabilities of an
  anisotropically expanding non-Abelian plasma: 1D+3V discretized hard-loop
  simulations}}, \href{https://doi.org/10.1103/PhysRevD.78.045023}{\emph{Phys.
  Rev.} {\bfseries D78} (2008) 045023}
  [\href{https://arxiv.org/abs/0802.1714}{{\ttfamily 0802.1714}}].

\bibitem{Fukushima:2011nq}
K.~Fukushima and F.~Gelis, \emph{{The evolving Glasma}}, {\emph{Nucl.Phys.}
  {\bfseries A874} (2012) 108}
  [\href{https://arxiv.org/abs/1106.1396}{{\ttfamily 1106.1396}}].

\bibitem{Kurkela:2011ti}
A.~Kurkela and G.~D. Moore, \emph{{Thermalization in Weakly Coupled Nonabelian
  Plasmas}}, {\emph{JHEP} {\bfseries 1112} (2011) 044}
  [\href{https://arxiv.org/abs/1107.5050}{{\ttfamily 1107.5050}}].

\bibitem{Kurkela:2011ub}
A.~Kurkela and G.~D. Moore, \emph{{Bjorken Flow, Plasma Instabilities, and
  Thermalization}}, {\emph{JHEP} {\bfseries 1111} (2011) 120}
  [\href{https://arxiv.org/abs/1108.4684}{{\ttfamily 1108.4684}}].

\bibitem{Blaizot:2011xf}
J.-P. Blaizot, F.~Gelis, J.-F. Liao, L.~McLerran and R.~Venugopalan,
  \emph{{Bose--Einstein Condensation and Thermalization of the Quark Gluon
  Plasma}},
  \href{https://doi.org/10.1016/j.nuclphysa.2011.10.005}{\emph{Nucl.Phys.}
  {\bfseries A873} (2012) 68}
  [\href{https://arxiv.org/abs/1107.5296}{{\ttfamily 1107.5296}}].

\bibitem{Attems:2012js}
M.~Attems, A.~Rebhan and M.~Strickland, \emph{{Instabilities of an
  anisotropically expanding non-Abelian plasma: 3D+3V discretized hard-loop
  simulations}},
  \href{https://doi.org/10.1103/PhysRevD.87.025010}{\emph{Phys.Rev.} {\bfseries
  D87} (2013) 025010} [\href{https://arxiv.org/abs/1207.5795}{{\ttfamily
  1207.5795}}].

\bibitem{Berges:2012iw}
J.~Berges, K.~Boguslavski and S.~Schlichting, \emph{{Nonlinear amplification of
  instabilities with longitudinal expansion}}, {\emph{Phys.Rev.} {\bfseries
  D85} (2012) 076005} [\href{https://arxiv.org/abs/1201.3582}{{\ttfamily
  1201.3582}}].

\bibitem{Gelis:2013rba}
T.~Epelbaum and F.~Gelis, \emph{{Pressure isotropization in high energy heavy
  ion collisions}},
  \href{https://doi.org/10.1103/PhysRevLett.111.232301}{\emph{Phys.Rev.Lett.}
  {\bfseries 111} (2013) 232301}
  [\href{https://arxiv.org/abs/1307.2214}{{\ttfamily 1307.2214}}].

\bibitem{Chesler:2008hg}
P.~M. Chesler and L.~G. Yaffe, \emph{{Horizon formation and
  far-from-equilibrium isotropization in supersymmetric Yang-Mills plasma}},
  \href{https://doi.org/10.1103/PhysRevLett.102.211601}{\emph{Phys.Rev.Lett.}
  {\bfseries 102} (2009) 211601}
  [\href{https://arxiv.org/abs/0812.2053}{{\ttfamily 0812.2053}}].

\bibitem{Beuf:2009cx}
G.~Beuf, M.~P. Heller, R.~A. Janik and R.~Peschanski, \emph{{Boost-invariant
  early time dynamics from AdS/CFT}},
  \href{https://doi.org/10.1088/1126-6708/2009/10/043}{\emph{JHEP} {\bfseries
  10} (2009) 043} [\href{https://arxiv.org/abs/0906.4423}{{\ttfamily
  0906.4423}}].

\bibitem{Chesler:2009cy}
P.~M. Chesler and L.~G. Yaffe, \emph{{Boost invariant flow, black hole
  formation, and far-from-equilibrium dynamics in N = 4 supersymmetric
  Yang-Mills theory}},
  \href{https://doi.org/10.1103/PhysRevD.82.026006}{\emph{Phys.Rev.} {\bfseries
  D82} (2010) 026006} [\href{https://arxiv.org/abs/0906.4426}{{\ttfamily
  0906.4426}}].

\bibitem{Heller:2011ju}
M.~P. Heller, R.~A. Janik and P.~Witaszczyk, \emph{{The characteristics of
  thermalization of boost-invariant plasma from holography}},
  {\emph{Phys.Rev.Lett.} {\bfseries 108} (2012) 201602}
  [\href{https://arxiv.org/abs/1103.3452}{{\ttfamily 1103.3452}}].

\bibitem{Heller:2012je}
M.~P. Heller, R.~A. Janik and P.~Witaszczyk, \emph{{A numerical relativity
  approach to the initial value problem in asymptotically Anti-de Sitter
  spacetime for plasma thermalization - an ADM formulation}}, {\emph{Phys.Rev.}
  {\bfseries D85} (2012) 126002}
  [\href{https://arxiv.org/abs/1203.0755}{{\ttfamily 1203.0755}}].

\bibitem{Heller:2012km}
M.~P. Heller, D.~Mateos, W.~van~der Schee and D.~Trancanelli, \emph{{Strong
  Coupling Isotropization of Non-Abelian Plasmas Simplified}},
  {\emph{Phys.Rev.Lett.} {\bfseries 108} (2012) 191601}
  [\href{https://arxiv.org/abs/1202.0981}{{\ttfamily 1202.0981}}].

\bibitem{vanderSchee:2012qj}
W.~van~der Schee, \emph{{Holographic thermalization with radial flow}},
  \href{https://doi.org/10.1103/PhysRevD.87.061901}{\emph{Phys.Rev.} {\bfseries
  D87} (2013) 061901} [\href{https://arxiv.org/abs/1211.2218}{{\ttfamily
  1211.2218}}].

\bibitem{Casalderrey-Solana:2013aba}
J.~Casalderrey-Solana, M.~P. Heller, D.~Mateos and W.~van~der Schee,
  \emph{{From full stopping to transparency in a holographic model of heavy ion
  collisions}},
  \href{https://doi.org/10.1103/PhysRevLett.111.181601}{\emph{Phys. Rev. Lett.
  111,} {\bfseries 181601} (2013) }
  [\href{https://arxiv.org/abs/1305.4919}{{\ttfamily 1305.4919}}].

\bibitem{Heller:2013oxa}
M.~P. Heller, D.~Mateos, W.~van~der Schee and M.~Triana, \emph{{Holographic
  isotropization linearized}},
  \href{https://doi.org/10.1007/JHEP09(2013)026}{\emph{JHEP} {\bfseries 09}
  (2013) 026} [\href{https://arxiv.org/abs/1304.5172}{{\ttfamily 1304.5172}}].

\bibitem{Keegan:2015avk}
L.~Keegan, A.~Kurkela, P.~Romatschke, W.~van~der Schee and Y.~Zhu, \emph{{Weak
  and strong coupling equilibration in nonabelian gauge theories}},
  \href{https://doi.org/10.1007/JHEP04(2016)031}{\emph{JHEP} {\bfseries 04}
  (2016) 031} [\href{https://arxiv.org/abs/1512.05347}{{\ttfamily
  1512.05347}}].

\bibitem{Chesler:2015bba}
P.~M. Chesler, \emph{{Colliding shock waves and hydrodynamics in small
  systems}}, \href{https://doi.org/10.1103/PhysRevLett.115.241602}{\emph{Phys.
  Rev. Lett.} {\bfseries 115} (2015) 241602}
  [\href{https://arxiv.org/abs/1506.02209}{{\ttfamily 1506.02209}}].

\bibitem{Kurkela:2015qoa}
A.~Kurkela and Y.~Zhu, \emph{{Isotropization and hydrodynamization in weakly
  coupled heavy-ion collisions}},
  \href{https://doi.org/10.1103/PhysRevLett.115.182301}{\emph{Phys. Rev. Lett.}
  {\bfseries 115} (2015) 182301}
  [\href{https://arxiv.org/abs/1506.06647}{{\ttfamily 1506.06647}}].

\bibitem{Chesler:2016ceu}
P.~M. Chesler, \emph{{How big are the smallest drops of quark-gluon plasma?}},
  \href{https://doi.org/10.1007/JHEP03(2016)146}{\emph{JHEP} {\bfseries 03}
  (2016) 146} [\href{https://arxiv.org/abs/1601.01583}{{\ttfamily
  1601.01583}}].

\bibitem{Attems:2016ugt}
M.~Attems, J.~Casalderrey-Solana, D.~Mateos, I.~Papadimitriou,
  D.~Santos-Oliv·n, C.~F. Sopuerta et~al., \emph{{Thermodynamics, transport
  and relaxation in non-conformal theories}},
  \href{https://doi.org/10.1007/JHEP10(2016)155}{\emph{JHEP} {\bfseries 10}
  (2016) 155} [\href{https://arxiv.org/abs/1603.01254}{{\ttfamily
  1603.01254}}].

\bibitem{Attems:2016tby}
M.~Attems, J.~Casalderrey-Solana, D.~Mateos, D.~Santos-Oliv·n, C.~F. Sopuerta,
  M.~Triana et~al., \emph{{Holographic Collisions in Non-conformal Theories}},
  \href{https://doi.org/10.1007/JHEP01(2017)026}{\emph{JHEP} {\bfseries 01}
  (2017) 026} [\href{https://arxiv.org/abs/1604.06439}{{\ttfamily
  1604.06439}}].

\bibitem{Attems:2017zam}
M.~Attems, J.~Casalderrey-Solana, D.~Mateos, D.~Santos-Oliv·n, C.~F. Sopuerta,
  M.~Triana et~al., \emph{{Paths to equilibrium in non-conformal collisions}},
  \href{https://doi.org/10.1007/JHEP06(2017)154}{\emph{JHEP} {\bfseries 06}
  (2017) 154} [\href{https://arxiv.org/abs/1703.09681}{{\ttfamily
  1703.09681}}].

\bibitem{Strickland:2013uga}
M.~Strickland, \emph{{Thermalization and isotropization in heavy-ion
  collisions}}, \href{https://doi.org/10.1007/s12043-015-0972-1}{\emph{Pramana}
  {\bfseries 84} (2015) 671} [\href{https://arxiv.org/abs/1312.2285}{{\ttfamily
  1312.2285}}].

\bibitem{Noronha:2011fi}
J.~Noronha and G.~S. Denicol, \emph{{Transient Fluid Dynamics of the
  Quark-Gluon Plasma According to AdS/CFT}},
  \href{https://arxiv.org/abs/1104.2415}{{\ttfamily 1104.2415}}.

\bibitem{Heller:2015dha}
M.~P. Heller and M.~Spalinski, \emph{{Hydrodynamics Beyond the Gradient
  Expansion: Resurgence and Resummation}},
  \href{https://doi.org/10.1103/PhysRevLett.115.072501}{\emph{Phys. Rev. Lett.}
  {\bfseries 115} (2015) 072501}
  [\href{https://arxiv.org/abs/1503.07514}{{\ttfamily 1503.07514}}].

\bibitem{Florkowski:2017olj}
W.~Florkowski, M.~P. Heller and M.~Spalinski, \emph{{New theories of
  relativistic hydrodynamics in the LHC era}},
  \href{https://arxiv.org/abs/1707.02282}{{\ttfamily 1707.02282}}.

\bibitem{Romatschke:2017vte}
P.~Romatschke, \emph{{Relativistic Fluid Dynamics Far From Local Equilibrium}},
  \href{https://doi.org/10.1103/PhysRevLett.120.012301}{\emph{Phys. Rev. Lett.}
  {\bfseries 120} (2018) 012301}
  [\href{https://arxiv.org/abs/1704.08699}{{\ttfamily 1704.08699}}].

\bibitem{Bemfica:2017wps}
F.~S. Bemfica, M.~M. Disconzi and J.~Noronha, \emph{{Causality and existence of
  solutions of relativistic viscous fluid dynamics with gravity}},
  \href{https://arxiv.org/abs/1708.06255}{{\ttfamily 1708.06255}}.

\bibitem{Spalinski:2017mel}
M.~Spalinski, \emph{{On the hydrodynamic attractor of Yang--Mills plasma}},
  \href{https://doi.org/10.1016/j.physletb.2017.11.059}{\emph{Phys. Lett.}
  {\bfseries B776} (2018) 468}
  [\href{https://arxiv.org/abs/1708.01921}{{\ttfamily 1708.01921}}].

\bibitem{Romatschke:2017acs}
P.~Romatschke, \emph{{Relativistic Hydrodynamic Attractors with Broken
  Symmetries: Non-Conformal and Non-Homogeneous}},
  \href{https://doi.org/10.1007/JHEP12(2017)079}{\emph{JHEP} {\bfseries 12}
  (2017) 079} [\href{https://arxiv.org/abs/1710.03234}{{\ttfamily
  1710.03234}}].

\bibitem{Behtash:2017wqg}
A.~Behtash, C.~N. Cruz-Camacho and M.~Martinez, \emph{{Far-from-equilibrium
  attractors and nonlinear dynamical systems approach to the Gubser flow}},
  \href{https://arxiv.org/abs/1711.01745}{{\ttfamily 1711.01745}}.

\bibitem{Florkowski:2017jnz}
W.~Florkowski, E.~Maksymiuk and R.~Ryblewski, \emph{{Coupled kinetic equations
  for fermions and bosons in the relaxation-time approximation}},
  \href{https://doi.org/10.1103/PhysRevC.97.024915}{\emph{Phys. Rev.}
  {\bfseries C97} (2018) 024915}
  [\href{https://arxiv.org/abs/1710.07095}{{\ttfamily 1710.07095}}].

\bibitem{Florkowski:2017ovw}
W.~Florkowski, E.~Maksymiuk and R.~Ryblewski, \emph{{Anisotropic-hydrodynamics
  approach to a quark-gluon fluid mixture}},
  \href{https://doi.org/10.1103/PhysRevC.97.014904}{\emph{Phys. Rev.}
  {\bfseries C97} (2018) 014904}
  [\href{https://arxiv.org/abs/1711.03872}{{\ttfamily 1711.03872}}].

\bibitem{Strickland:2017kux}
M.~Strickland, J.~Noronha and G.~Denicol, \emph{{The anisotropic
  non-equilibrium hydrodynamic attractor}},
  \href{https://arxiv.org/abs/1709.06644}{{\ttfamily 1709.06644}}.

\bibitem{Almaalol:2018ynx}
D.~Almaalol and M.~Strickland, \emph{{Anisotropic hydrodynamics with a scalar
  collisional kernel}},
  \href{https://doi.org/10.1103/PhysRevC.97.044911}{\emph{Phys. Rev.}
  {\bfseries C97} (2018) 044911}
  [\href{https://arxiv.org/abs/1801.10173}{{\ttfamily 1801.10173}}].

\bibitem{Denicol:2018pak}
G.~S. Denicol and J.~Noronha, \emph{{Hydrodynamic attractor and the fate of
  perturbative expansions in Gubser flow}},
  \href{https://arxiv.org/abs/1804.04771}{{\ttfamily 1804.04771}}.

\bibitem{Behtash:2018moe}
A.~Behtash, S.~Kamata, M.~Martinez and C.~N. Cruz-Camacho,
  \emph{{Non-perturbative rheological behavior of a far-from-equilibrium
  expanding plasma}},  \href{https://arxiv.org/abs/1805.07881}{{\ttfamily
  1805.07881}}.

\bibitem{Florkowski:2013lza}
W.~Florkowski, R.~Ryblewski and M.~Strickland, \emph{{Anisotropic Hydrodynamics
  for Rapidly Expanding Systems}},
  \href{https://arxiv.org/abs/1304.0665}{{\ttfamily 1304.0665}}.

\bibitem{Florkowski:2013lya}
W.~Florkowski, R.~Ryblewski and M.~Strickland, \emph{{Testing viscous and
  anisotropic hydrodynamics in an exactly solvable case}},
  \href{https://arxiv.org/abs/1305.7234}{{\ttfamily 1305.7234}}.

\bibitem{Florkowski:2014sfa}
W.~Florkowski, E.~Maksymiuk, R.~Ryblewski and M.~Strickland, \emph{{Exact
  solution of the (0+1)-dimensional Boltzmann equation for a massive gas}},
  \href{https://doi.org/10.1103/PhysRevC.89.054908}{\emph{Phys. Rev.}
  {\bfseries C89} (2014) 054908}
  [\href{https://arxiv.org/abs/1402.7348}{{\ttfamily 1402.7348}}].

\bibitem{Florkowski:2014sda}
W.~Florkowski and E.~Maksymiuk, \emph{{Exact solution of the (0+1)-dimensional
  Boltzmann equation for massive Bose-Einstein and Fermi-Dirac gases}},
  \href{https://doi.org/10.1088/0954-3899/42/4/045106}{\emph{J. Phys.}
  {\bfseries G42} (2015) 045106}
  [\href{https://arxiv.org/abs/1411.3666}{{\ttfamily 1411.3666}}].

\bibitem{Denicol:2014xca}
G.~S. Denicol, U.~W. Heinz, M.~Martinez, J.~Noronha and M.~Strickland,
  \emph{{New Exact Solution of the Relativistic Boltzmann Equation and its
  Hydrodynamic Limit}},
  \href{https://doi.org/10.1103/PhysRevLett.113.202301}{\emph{Phys. Rev. Lett.}
  {\bfseries 113} (2014) 202301}
  [\href{https://arxiv.org/abs/1408.5646}{{\ttfamily 1408.5646}}].

\bibitem{Denicol:2014tha}
G.~S. Denicol, U.~W. Heinz, M.~Martinez, J.~Noronha and M.~Strickland,
  \emph{{Studying the validity of relativistic hydrodynamics with a new exact
  solution of the Boltzmann equation}},
  \href{https://doi.org/10.1103/PhysRevD.90.125026}{\emph{Phys. Rev.}
  {\bfseries D90} (2014) 125026}
  [\href{https://arxiv.org/abs/1408.7048}{{\ttfamily 1408.7048}}].

\bibitem{Maksymiuk:2017cnv}
E.~Maksymiuk, \emph{{Kinetic equations and anisotropic hydrodynamics for quark
  and gluon fluids}},  \href{https://arxiv.org/abs/1712.01591}{{\ttfamily
  1712.01591}}.

\bibitem{Denicol:2010xn}
G.~S. Denicol, T.~Koide and D.~H. Rischke, \emph{{Dissipative relativistic
  fluid dynamics: a new way to derive the equations of motion from kinetic
  theory}}, \href{https://doi.org/10.1103/PhysRevLett.105.162501}{\emph{Phys.
  Rev. Lett.} {\bfseries 105} (2010) 162501}
  [\href{https://arxiv.org/abs/1004.5013}{{\ttfamily 1004.5013}}].

\bibitem{Denicol:2011fa}
G.~S. Denicol, J.~Noronha, H.~Niemi and D.~H. Rischke, \emph{{Origin of the
  Relaxation Time in Dissipative Fluid Dynamics}},
  \href{https://doi.org/10.1103/PhysRevD.83.074019}{\emph{Phys. Rev.}
  {\bfseries D83} (2011) 074019}
  [\href{https://arxiv.org/abs/1102.4780}{{\ttfamily 1102.4780}}].

\bibitem{Jaiswal:2013npa}
A.~Jaiswal, \emph{{Relativistic dissipative hydrodynamics from kinetic theory
  with relaxation time approximation}},
  \href{https://doi.org/10.1103/PhysRevC.87.051901}{\emph{Phys. Rev.}
  {\bfseries C87} (2013) 051901}
  [\href{https://arxiv.org/abs/1302.6311}{{\ttfamily 1302.6311}}].

\bibitem{Jaiswal:2013vta}
A.~Jaiswal, \emph{{Relativistic third-order dissipative fluid dynamics from
  kinetic theory}},
  \href{https://doi.org/10.1103/PhysRevC.88.021903}{\emph{Phys. Rev.}
  {\bfseries C88} (2013) 021903}
  [\href{https://arxiv.org/abs/1305.3480}{{\ttfamily 1305.3480}}].

\bibitem{Denicol:2014mca}
G.~S. Denicol, W.~Florkowski, R.~Ryblewski and M.~Strickland, \emph{{Shear-bulk
  coupling in nonconformal hydrodynamics}},
  \href{https://doi.org/10.1103/PhysRevC.90.044905}{\emph{Phys.Rev.} {\bfseries
  C90} (2014) 044905} [\href{https://arxiv.org/abs/1407.4767}{{\ttfamily
  1407.4767}}].

\bibitem{Florkowski:2015lra}
W.~Florkowski, A.~Jaiswal, E.~Maksymiuk, R.~Ryblewski and M.~Strickland,
  \emph{{Relativistic quantum transport coefficients for second-order viscous
  hydrodynamics}},
  \href{https://doi.org/10.1103/PhysRevC.91.054907}{\emph{Phys. Rev.}
  {\bfseries C91} (2015) 054907}
  [\href{https://arxiv.org/abs/1503.03226}{{\ttfamily 1503.03226}}].

\bibitem{Bialas:1984wv}
A.~Bia\l{}as and W.~Czy\ifmmode~\dot{z}\else \.{z}\fi{}, \emph{Boost-invariant
  boltzmann-vlasov equations for relativistic quark-antiquark plasma},
  \href{https://doi.org/10.1103/PhysRevD.30.2371}{\emph{Phys. Rev. D}
  {\bfseries 30} (1984) 2371}.

\bibitem{Bialas:1987en}
A.~Bia\l{}as and W.~Czy\ifmmode~\dot{z}\else \.{z}\fi{}, \emph{Oscillations of
  quark-gluon plasma generated in strong color fields},
  \href{https://doi.org/10.1016/0550-3213(88)90035-1}{\emph{Nuclear Physics B}
  {\bfseries 296} (1988) 611 }.

\bibitem{Baym:1984np}
G.~Baym, \emph{{Thermal equilibration in Ultrarelativistic Heavy Ion
  Collisions}}, {\emph{Phys. Lett.} {\bfseries B138} (1984) 18}.

\bibitem{Florkowski:2010cf}
W.~Florkowski and R.~Ryblewski, \emph{{Highly-anisotropic and
  strongly-dissipative hydrodynamics for early stages of relativistic heavy-ion
  collisions}}, \href{https://doi.org/10.1103/PhysRevC.83.034907}{\emph{Phys.
  Rev.} {\bfseries C83} (2011) 034907}
  [\href{https://arxiv.org/abs/1007.0130}{{\ttfamily 1007.0130}}].

\bibitem{Martinez:2010sc}
M.~Martinez and M.~Strickland, \emph{{Dissipative Dynamics of Highly
  Anisotropic Systems}},
  \href{https://doi.org/10.1016/j.nuclphysa.2010.08.011}{\emph{Nucl. Phys.}
  {\bfseries A848} (2010) 183}
  [\href{https://arxiv.org/abs/1007.0889}{{\ttfamily 1007.0889}}].

\bibitem{Tinti:2013vba}
L.~Tinti and W.~Florkowski, \emph{{Projection method and new formulation of
  leading-order anisotropic hydrodynamics}},
  \href{https://doi.org/10.1103/PhysRevC.89.034907}{\emph{Phys.Rev.} {\bfseries
  C89} (2014) 034907} [\href{https://arxiv.org/abs/1312.6614}{{\ttfamily
  1312.6614}}].

\bibitem{Alqahtani:2017mhy}
M.~Alqahtani, M.~Nopoush and M.~Strickland, \emph{{Relativistic anisotropic
  hydrodynamics}},
  \href{https://doi.org/10.1016/j.ppnp.2018.05.004}{\emph{Prog. Part. Nucl.
  Phys.} {\bfseries 101} (2018) 204}
  [\href{https://arxiv.org/abs/1712.03282}{{\ttfamily 1712.03282}}].

\bibitem{MikeCodeDB}
M.~Strickland. \url{http://personal.kent.edu/~mstrick6/code/}, 2017.

\bibitem{Heller:2016rtz}
M.~P. Heller, A.~Kurkela, M.~Spali\'nski and V.~Svensson,
  \emph{{Hydrodynamization in kinetic theory: Transient modes and the gradient
  expansion}}, \href{https://doi.org/10.1103/PhysRevD.97.091503}{\emph{Phys.
  Rev. D} {\bfseries 97} (2018) 091503}
  [\href{https://arxiv.org/abs/1609.04803}{{\ttfamily 1609.04803}}].

\bibitem{Jaiswal:2021uvv}
S.~Jaiswal, C.~Chattopadhyay, L.~Du, U.~Heinz and S.~Pal, \emph{{On
  non-conformal kinetic theory and hydrodynamics for Bjorken flow}},
  \href{https://arxiv.org/abs/2107.10248}{{\ttfamily 2107.10248}}.

\end{thebibliography}\endgroup

\end{document}